\documentclass[smallextended,final]{svjour3}	

\smartqed  

\usepackage{graphicx}

\usepackage{amssymb,amsmath} 
\usepackage{cite}
\usepackage{graphics}
\usepackage{subfigure}
\usepackage{epsfig}
\usepackage{algorithm}
\usepackage{algpseudocode}

\begin{document}

\title{A population Monte Carlo scheme with transformed weights and its application
to stochastic kinetic models
}

\titlerunning{Population Monte Carlo with transformed weights}        

\author{Eugenia Koblents         \and
        Joaqu\'{\i}n M\'{\i}guez }


\institute{E. Koblents and J. M\'{\i}guez \at
              Department of Signal Theory and Communications, Universidad Carlos III de Madrid, Madrid, Spain \\
              \email{ekoblents,jmiguez@tsc.uc3m.es}           
}

\date{Received: date / Accepted: date}

\maketitle

\begin{abstract}

This paper addresses the problem of Monte Carlo approximation of
posterior probability distributions. In particular, we have considered a
recently proposed technique known as population Monte Carlo (PMC),
which is based on an iterative importance sampling approach. An
important drawback of this methodology is the degeneracy of the 
importance weights when the dimension of either the observations or 
the variables of interest is high. To alleviate this difficulty, we propose 
a novel method that performs a nonlinear transformation on the 
importance weights. This operation reduces the weight variation, hence
it avoids their degeneracy and increases the efficiency of the importance
sampling scheme, specially when drawing from a proposal functions
which are poorly adapted to the true posterior.

For the sake of illustration, we have applied the proposed algorithm to the 
estimation of the parameters of a Gaussian mixture model. This is a very 
simple problem that enables us to clearly show and discuss the main features 
of the proposed technique. As a practical application, we have also 
considered the popular (and challenging) problem of estimating the rate
parameters of stochastic kinetic models (SKM). SKMs are highly multivariate 
systems that model molecular interactions in biological and chemical problems. 
We introduce a particularization of the proposed algorithm to SKMs and 
present numerical results. 

%
%


\end{abstract}

\section{Introduction}
\label{Introduction}


The problem of performing inference in high dimensional spaces
appears in many practical applications. For example, it is of
increasing interest in the biological sciences to develop new
techniques that allow for the efficient estimation of the parameters
governing the behavior of complex autoregulatory networks. The main
difficulty often encountered when tackling this kind of problems is
the design of numerical inference algorithms which are stable and
have a guaranteed convergence also in high-dimensional spaces.


A very common strategy, which has been successfully applied in a
broad variety of complex problems, is the Monte Carlo methodology.
In particular, we have considered a recently proposed technique
known as population Monte Carlo (PMC) \cite{Cappe04}, which is based
on an iterative importance sampling approach. The aim of this method
is the approximation of static probability distributions by way of
discrete random measures consisting of samples and associated weights. The
target distribution is often the posterior distribution of a set
variables of interest, given some observed data.

The main advantages of the PMC scheme, compared to the widely
established Markov chain Monte Carlo (MCMC) methodology, 
are the possibility of developing parallel implementations, the sample 
independence and the fact that an unbiased estimate is provided at 
each iteration, which avoids the need of a convergence period.


On the contrary, an important drawback of the importance sampling
approach, and particulary of PMC, is that its performance heavily
depends on the choice of the proposal distribution (or importance
function) that is used to generate the samples and compute the
weights. When the dimension of the variables of interest is large, or
the target probability density function (pdf) is very sharp with respect 
to the proposal (this occurs when, e.g., the number of observations
is high or when the prior distribution is uninformative), the importance 
weights degenerate\footnote{In this context, degeneracy means that
the vast majority of importance weights become negligible (practically
zero) except for a very small number of them 
\cite{Kong94,Doucet00}.}  
leading to an extremely low number of representative samples. This
problem is commonly known as weight degeneracy and is closely related 
to the ``curse of dimensionality''. To the best of our knowledge, 
the degeneracy problem has not been successfully addressed so far in 
the PMC framework. The issue was already mentioned in the original paper 
\cite{Cappe04}, though, noting that the proposed scheme did not provide
neither a stabilization of the so called effective sample size, nor
of the variance of the importance weights.

The effort in the field of PMC algorithms has been mainly directed toward
the design of efficient proposal functions. A recently proposed scheme for 
the proposal update is the mixture PMC technique \cite{Cappe08}, that 
models the importance functions as mixtures of kernels. Both the weights 
and the internal parameters of each mixture component are adapted along 
the iterations with the goal of minimizing the Kullback-Leiber divergence 
between the target density and the proposal. However, this scheme also 
suffers from degeneracy unless applied to very simple examples and the 
authors of \cite{Cappe08} propose to apply an additional Rao-Blackwellization 
scheme to mitigate the problems that appear in multidimensional problems.

Another recently proposed PMC scheme  \cite{Djuric11} is based on the
Gibbs sampling method and constructs the importance functions as
alternating conditional densities. Thus, this method allows to sample efficiently 
from high-dimensional proposals. However, the importance weights still 
present severe degeneracy due to the extreme values of the likelihood 
function in high-dimensional spaces, unless the number of samples is 
extremely high. The technique is based on the multiple marginalized PMC, 
proposed in \cite{Bugallo09}, \cite{Shen10}.




In this paper we propose a novel PMC method, termed nonlinear PMC (NPMC). 
The emphasis is not placed on the proposal update scheme, which can be very
simple (here we restrict ourselves to importance functions chosen as multivariate 
normal denisities whose parameters are adjusted along the iterations to match the
moments of the latest approximation of the posterior distribution). The main feature
of the technique is the application of a nonlinear transformation to the importance 
weights in order to reduce their variations. In this way, the efficiency of the sampling 
scheme is improved (specially when drawing from ``poor'' proposals) and, most
importantly, the degeneracy of the weights is drastically mitigated even when the 
number of generated samples is relatively small. We show analytically that the 
approximate integrals computed using the transformed weights converge 
asymptotically to the true integrals.

In this work we have successfully applied the proposed methodology
to two different problems where severe weight
degeneracy is observed. The first example is a Gaussian mixture
model, already discussed in \cite{Cappe04}. We have used this example to
illustrate the degeneracy problem and for comparison of the original
PMC scheme with the proposed methods. We provide numerical results
that suggest that degeneracy of the importance weights takes places
even in simple and low-dimensional problems, when the proposal pdf
is wide with respect to the target. The proposed method clearly
outperforms the original PMC scheme and provides a high number of
representative samples and very accurate estimates of the variables
of interest.

As a practical application, we have also applied the
proposed algorithm to the popular (and challenging) problem
of estimating the rate parameters in stochastic kinetic
models \cite{Boys08}. Such models describe the time evolution of the
population of a set of species or chemical reactions, which evolve
according to a set of constant rate parameters, and present an
autoregulatory behavior. This problem is currently of great interest
in a broad diversity of biological and molecular problems, such as
complex auto-regulatory gene networks.

We introduce a particularization of the generic NPMC algorithm to
SKMs, which tackles the evaluation of the likelihood function (which 
can not be computed exactly) using a particle filter. As a simple and intuitive,
yet physically meaningful example, we have obtained numerical
results for the Lotka-Volterra model, also known as predator-prey
model, consisting of two interacting species related by three
reactions with associated unknown rates. The proposed method
provides a very good performance also in this scenario, where 
standard PMC algorithms can be easily shown to fail.

%


The rest of the paper is organized as follows. In Section
\ref{Notation} we introduce some notation. In Section \ref{ProbStat}, we give a formal statement of
the class of problems addressed in this paper. In Section
\ref{PMC_section}, the population Monte Carlo algorithm is presented
and a discussion of the weight degeneracy problem is
provided. In section \ref{Algorithms} the proposed algorithm is
described. In section \ref{Analysis} we provide a
theoretical convergence analysis of the new methodology. In
section \ref{GMM_example} we present numerical results on a Gaussian
mixture model. We illustrate the effects of degeneracy and the
performance of the proposed algorithms in this scenario. In Section
\ref{SKM_example}, we describe the practical application of the
proposed algorithm to the problem of estimating the constant rates of a
stochastic kinetic model, and present numerical results. Section
\ref{Conclusion} is devoted to the conclusions.


\section{Notation}
\label{Notation}

In this section we introduce some notations that are used through
the paper.

\begin{itemize}

\item All vectors are assumed to have column form. We denote them using boldface lower-case letters, e.g., $\boldsymbol{\theta}$, $\textbf{y}$, and its $k$-th scalar component is written as the corresponding normal-face letter with a subscript, e.g., $\theta_k$, $y_k$. Matrices are denoted by boldface upper-case letters, e.g., $\boldsymbol{\Sigma}$. 
$N$.

\item We use $\mathbb{R}^K$, with integer $K \geq 1$, to denote the
set of $K$-dimensional vectors with real entries.

\item $B(\mathbb{R}^K)$ is the set of bounded real functions over $\mathbb{R}^K$. In particular, the supremum norm of a real function $f: \mathbb{R}^K \rightarrow \mathbb{R}$ is denoted as $\| f \|_\infty = \sup_{\textbf{z}\in\mathbb{R}^K} | f(\textbf{z}) |$ and $f \in B(\mathbb{R}^K)$ when $\| f \|_\infty < \infty$.

\item The target pdf of a PMC algorithm is denoted as $\pi$, the proposal density as $q$ and the rest of pdf's as $p$.

\item We write conditional pdf's as $p(\textbf{y} | \boldsymbol{\theta})$,
and joint densities as $p(\boldsymbol{\theta}) = p(\theta_1, \ldots, \theta_K)$. This is an argument-wise notation, hence
$p(\theta_1)$ denotes the distribution of $\theta_1$, possibly
different from $p(\theta_2)$, which represents the distribution of
$\theta_2$.

\item The integral of a function $f$ with respect to a measure $\mu$ is denoted by the shorthand $(f,\mu) = \int f(\boldsymbol{\theta})\mu(d\boldsymbol{\theta})$. If the measure has a density $\pi$, i.e., $\mu(d\boldsymbol{\theta})=\pi(\boldsymbol{\theta})d\boldsymbol{\theta}$, then we also write $(f,\pi)=\int f(\boldsymbol{\theta}) \pi(\boldsymbol{\theta})d\boldsymbol{\theta}$.


\item $E_{p(\boldsymbol{\theta})}[f(\boldsymbol{\theta})]$ denotes expectation of the function $f$ with respect to the probability distribution with density $p(\boldsymbol{\theta})$.


\item A sample from the distribution of the random vector $\boldsymbol{\theta}$ is denoted by $\boldsymbol{\theta}^{(i)}$.
Sets of $M$ samples $\{ \boldsymbol{\theta}^{(1)}, \ldots,
\boldsymbol{\theta}^{(M)} \}$ are denoted as $\{
\boldsymbol{\theta}^{(i)} \}_{i=1}^M$.

\item $\delta_{\theta^{(i)}}(d\boldsymbol{\theta})$ is the unit delta measure located at $\boldsymbol{\theta} = \boldsymbol{\theta}^{(i)}$.

\item Unweighted samples are denoted with a tilde as $\left\{ \tilde{\boldsymbol{\theta}}^{(i)} \right\}$, opposite to the corresponding
weighted samples $\left\{ \boldsymbol{\theta}^{(i)} \right\}$.



\item For a random value $X$, $\mathbb{P}\{ X < a \}$ denotes the probability of the event $X < a$.

\item In the chemical representation the notation $\alpha x$ must not be interpreted as a product, but rather is read as ``$\alpha$
molecules of species $x$''.


\end{itemize}

\section{Problem Statement}
\label{ProbStat}

Let $\boldsymbol{\theta} = \left[ \theta_1, \ldots, \theta_K
\right]^\top$ be a vector of $K$ unobserved real random variables
with prior density $p (\boldsymbol{\theta})$ and let $\textbf{y} =
\left[ y_1, \ldots, y_N \right]^\top$ be a vector of $N$ real random
observations related to $\boldsymbol{\theta}$ by way of a
conditional pdf $p(\textbf{y} |
\boldsymbol{\theta})$.

In this paper we address the problem of approximating the posterior
probability distribution of $\boldsymbol{\theta}$, i.e., the distribution with density $p(\boldsymbol{\theta} |
\textbf{y})$, using a random grid of $M$ points, $\{
\boldsymbol{\theta}^{(i)} \}_{i=1}^M$, in the space of the random
vector $\boldsymbol{\theta}$. Once the grid is generated, it is
simple to approximate any moments of $p(\boldsymbol{\theta} |
\textbf{y})$. For example, the posterior mean of $\boldsymbol{\theta}$ given the observations $\textbf{y}$ may be approximated as
\begin{equation*}
E_{p(\boldsymbol{\theta} | \textbf{y})} [\boldsymbol{\theta}] =
\int \boldsymbol{\theta} p(\boldsymbol{\theta}|\textbf{y})d\boldsymbol{\theta} \approx \frac{1}{M} \sum_{i=1}^M \boldsymbol{\theta}^{(i)}.
\end{equation*}

Unfortunately, the generation of useful samples that represent the probability measure
$p(\boldsymbol{\theta} | \textbf{y})d\boldsymbol{\theta}$ adequately when $K$ (or $N$)
is large is normally a very difficult task. The main goal of this work is to devise and assess an efficient
computational inference (Monte Carlo) methodology for the approximation of $p(\boldsymbol{\theta} | \textbf{y})d\boldsymbol{\theta}$ and its moments, i.e., expectations of the form $E_{p(\boldsymbol{\theta} | \textbf{y})}[ f(\boldsymbol{\theta} | \textbf{y}) ]$, where $f: \mathbb{R}^K \rightarrow \mathbb{R}$ is some integrable function of $\boldsymbol{\theta}$.


\section{Population Monte Carlo}
\label{PMC_section}

\subsection{Importance Sampling}

One of the main applications of statistical Monte Carlo methods is
the approximation of integrals of the form
\begin{equation*}
(f,\pi) = \int f(\boldsymbol{\theta}) \pi(\boldsymbol{\theta})
d\boldsymbol{\theta},
\end{equation*}
where $f$ is a real, integrable function of $\boldsymbol{\theta}$
and $\pi(\boldsymbol{\theta})$ is some pdf of interest (often termed
the \textit{target} density).
Given the random sample $\left\{ \boldsymbol{\theta}^{(i)}
\right\}_{i=1}^M$ we build a random discrete measure
\begin{equation*}
\pi^M(d \boldsymbol{\theta}) = \frac{1}{M} \sum_{i=1}^M
\delta_{\boldsymbol{\theta}^{(i)}}(d\boldsymbol{\theta})
\end{equation*}
that enables a straightforward approximation $(f,\pi^M)$ of
$(f,\pi)$, namely
\begin{equation*}
(f,\pi) \approx (f,\pi^M) = \int f(\boldsymbol{\theta})
\pi^M(d\boldsymbol{\theta}) = \frac{1}{M} \sum_{i=1}^M f \left(
\boldsymbol{\theta}^{(i)} \right).
\end{equation*}
Under mild assumptions, it can be shown that \cite{Robert04}
\begin{equation*}
\lim_{M\rightarrow \infty} (f,\pi^M) = (f,\pi) \quad \mbox{a.s.}
\end{equation*}

However, in many practical cases it is not possible to sample from
$\pi(\boldsymbol{\theta})$ directly. A common approach to overcome
this difficulty is to apply an importance sampling (IS) procedure
\cite{Robert04}. The key idea is to draw the samples $\{
\boldsymbol{\theta}^{(i)} \}_{i=1}^M$ from a (simpler) proposal pdf,
or importance function, $q(\boldsymbol{\theta})$, and then compute
associated importance weights of the form
\begin{equation*}
w^{(i)*} \propto \frac{\pi \left( \boldsymbol{\theta}^{(i)}
\right)}{q \left( \boldsymbol{\theta}^{(i)} \right)},
\end{equation*}
which we subsequently normalize to yield
\begin{equation*}
w^{(i)} = \frac{w^{(i)*}}{\sum_{j=1}^M w^{(j)*}},
\qquad i = 1, \ldots, M.
\end{equation*}
The integral $(f,\pi)$ is then approximated by the weighted sum
\begin{equation*}
(f,\pi^M) = \sum_{i=1}^M w^{(i)} f \left( \boldsymbol{\theta}^{(i)}
\right).
\end{equation*}


The efficiency of an IS algorithm depends heavily on the choice of
the proposal, $q(\boldsymbol{\theta})$. However, in order to ensure
the asymptotic convergence of the approximation $(f,\pi^M)$, when
$M$ is large enough, it is sufficient to select
$q(\boldsymbol{\theta})$ such that $q(\boldsymbol{\theta}) > 0$
whenever $\pi(\boldsymbol{\theta})
> 0$ \cite{Robert04}.

Finally, note that the computation of the normalized weights
requires that both $\pi(\boldsymbol{\theta})$ and
$q(\boldsymbol{\theta})$ can be evaluated up to a proportionality
constant independent of $\boldsymbol{\theta}$. This requirement is a mild one, since in most problems it
is possible to evaluate some function $h(\boldsymbol{\theta})
\propto \pi(\boldsymbol{\theta})$.

\subsection{Population Monte Carlo Algorithm}
\label{PMC_alg}

The population Monte Carlo (PMC) method \cite{Cappe04} is an
iterative IS scheme that seeks to generate a sequence of proposal
pdf's $q_\ell(\boldsymbol{\theta})$, $\ell = 1, \ldots, L$, such
that every new proposal is closer (in some adequate sense to be
defined) to the target density $\pi(\boldsymbol{\theta})$ than the
previous importance function. Such scheme demands, therefore, the
ability to learn about the target $\pi(\boldsymbol{\theta})$, given
the set of samples and weights at the $(\ell-1)$-th iteration, in
order to produce the new proposal $q_\ell(\boldsymbol{\theta})$ for
the $\ell$-th iteration. Taking this ability for granted, the
algorithm is simple and can be outlined as shown in Table \ref{tPMCAlgorithm}.

\vspace{0.3cm}

\begin{table}
\caption{Generic PMC algorithm}
\vspace{0.2cm}
\underline{\textbf{Iteration ($\ell = 0, \ldots, L$):}}

\begin{enumerate}
\item Select a proposal pdf $q_\ell(\boldsymbol{\theta})$.

\begin{itemize}
\item If $\pi(\boldsymbol{\theta})\propto p(\boldsymbol{\theta}|\textbf{y}) p(\boldsymbol{\theta})$, at iteration $\ell=0$ the proposal may be selected as the prior
$q_0(\boldsymbol{\theta}) = p(\boldsymbol{\theta})$.

\item At iterations $\ell = 1, \ldots, L$, the proposal pdf
$q_\ell(\boldsymbol{\theta})$ must be adapted according to the
weighted sample $\left\{ \boldsymbol{\theta}_{\ell-1}^{(i)},
w_{\ell-1}^{(i)} \right\}_{i=1}^M$ at the previous iteration.
\end{itemize}

\item Draw a collection of $M$ i.i.d. (independent and identically distributed) samples
$\Theta_\ell^M = \left\{ \boldsymbol{\theta}_\ell^{(i)}
\right\}_{i=1}^M$ from $q_\ell(\boldsymbol{\theta})$.

\item Compute the normalized weights
\begin{equation*}
w_\ell^{(i)} \propto \frac{\pi \left(
\boldsymbol{\theta}_\ell^{(i)} \right)}{q_\ell \left(
\boldsymbol{\theta}_\ell^{(i)} \right)}, \quad i = 1, \ldots, M.
\end{equation*}


\item Perform a resampling step according to the weights $w_\ell^{(i)}$ to create an unweighted sample set $\tilde{\Theta}_\ell^M = \left\{
\tilde{\boldsymbol{\theta}}_\ell^{(i)} \right\}_{i=1}^M$.
\end{enumerate}
\vspace{0.2cm}
\label{tPMCAlgorithm}
\end{table}

At every iteration of the algorithm it is possible to compute, if
needed, an estimate $(f,\pi_\ell^M)$ of $(f,\pi)$ as
\begin{equation*}
(f,\pi_\ell^M) = \sum_{i=1}^M w_\ell^{(i)} f \left(
\boldsymbol{\theta}_\ell^{(i)} \right)
\end{equation*}
and, if the proposals $q_\ell(\boldsymbol{\theta})$ are actually
improved across iterations, then it can be expected that the error
$
\left| (f,\pi) - (f,\pi_\ell^M) \right|
$
also decreases with $\ell$.

In problems of the type described in Section \ref{ProbStat}, the
target density is the posterior pdf of $\boldsymbol{\theta}$, i.e.,
$\pi(\boldsymbol{\theta}) = p(\boldsymbol{\theta} | \textbf{y})
\propto p(\textbf{y} | \boldsymbol{\theta})p(\boldsymbol{\theta})$,
where $p(\textbf{y} | \boldsymbol{\theta})$ is the likelihood
function of the variable $\boldsymbol{\theta}$ given the
observations $\textbf{y}$. A straightforward way of initializing the
algorithm is to use the prior as the starting proposal, $q_0
(\boldsymbol{\theta}) = p(\boldsymbol{\theta})$.

A frequently used index for the performance of Monte Carlo
sampling methods is the effective sample size
(ESS) $M^{eff}$ and its normalized version (NESS) $M^{neff}$,
respectively defined as \cite{Robert04}
\begin{equation*}
M^{eff} = \frac{1}{\sum_{i=1}^M \left( w^{(i)} \right) ^2}, \quad \mbox{and} \quad
M^{neff} = \frac{M^{eff}}{M}.
\end{equation*}
These indices may be used to quantitatively monitor the convergence
of the PMC algorithm and to stop the adaptation when the proposal
has converged to the target. Thus, ideally we expect to observe an
increase in both measures along the iterations, with $M^{neff}$
approaching unity as the algorithm converges.

However, unless the proposal pdf is well tailored to the target density, the
resulting importance weights will often present very large variations, leading
to a low number of effective samples. This problem is well known to
affect IS schemes and is usually termed as the degeneracy of the
weights \cite{Kong94,Doucet00}.

\subsection{Degeneracy of the importance weights}
\label{Degeneracy}

The degeneracy of the importance weights is a problem that arises when the
normalized importance weights $w^{(i)}$, $i=1, ..., M$, of a set of samples $\{
\boldsymbol{\theta}^{(i)} \}_{i=1}^M$ present large fluctuations and
the maximum of the weights, $\max_i w^{(i)}$, is close to one, leading
to an extremely low number of representative samples (i.e., samples with non
negligible weights). This situation occurs when the target and the
proposal densities are approximately mutually singular, i.e., they have
(essentially) have disjoint support. This problem has been addressed
in the scenario of very large scale systems \cite{Bengtsson08} and
it has been proved that the maximum of the sample weights converges
to one if the sample size $M$ is sub-exponential in the system
dimension $K$.

The degeneracy of the weights as the dimension of the
system increases is an intuitive fact and has been widely accepted
as one of the main drawbacks of IS schemes. Indeed, The degeneracy of the weights makes the application of IS to high dimensional systems (those with a large value of $K$) computationally intractable. However, not only a large value of $K$ leads to degeneracy. Indeed, it can be easily verified (numerically) that existing PMC methods can suffer from degeneracy even when applied to very simple (low dimensional) systems. Assume that the target pdf is the posterior, $\pi(\boldsymbol{\theta}) \propto p(\boldsymbol{\theta}|\textbf{y})p(\boldsymbol{\theta})$, and consider a set of $M$ samples $\{ \boldsymbol{\theta}^{(i)}
\}_{i=1}^M$ drawn from the prior pdf $p(\boldsymbol{\theta})$, which
is the case at the first iteration of the PMC algorithm. Assuming
conditionally independent observations, the importance weight
associated to the $i$-th sample is given by
\begin{equation}
w^{(i)} \propto p \left( \textbf{y} | \boldsymbol{\theta}^{(i)}
\right) = \prod_{n=1}^N p \left( y_n | \boldsymbol{\theta}^{(i)} \right), \quad i=1, \ldots,M.
\label{eqSharpDensity}
\end{equation}
Thus, the importance weights are obtained from a likelihood
consisting of the product of a potentially large number of factors. This
structure can lead to large fluctuations in the normalized importance
weights and a very low number of effective samples. In fact, the
degeneracy problem was already identified in the original work introducing the PMC methodology. The specific algorithm for which simulations are displayed in \cite{Cappe04} can be shown to present sharp variations in the effective sample size and a large variance of the estimators as well. However, to the best of our knowledge, no systematic solution has been provided so far for this problem.

A parallelism exists between high dimension and high number of
observations, both leading to computational problems. On one hand, as the
dimension $K$ increases, clearly the chance to obtain a
representative sample $\boldsymbol{\theta}^{(i)}$ decreases, since
the state space is larger. On the other hand, as the number of
observations increases, the probability concentrates in a smaller
region (the posterior pdf is sharper as a consequence, e.g., of a structure such as in Eq. \ref{eqSharpDensity}), which again leads to a low probability of obtaining representative samples.

Therefore, while the degeneracy of the weights increases critically with $K$ \cite{Bengtsson08}, in low dimensinal systems it is mainly motivated by a high number of observations $N$, unless the computational inference method is explicitly designed to account for this difficulty. In section \ref{GMM_example} we present numerical results to support this claim, which provides a rationale to understand the poor performance of existing PMC algorithms in certain low dimensional models.

In this paper we introduce a new methodology to tackle the weight degeneracy problem, either due to large $K$ or to large $N$. The key feature of the method is the application of a nonlinear transformation to the importance weights,
in order to reduce their variations and avoid degeneracy. As a
result, we obtain a number of effective samples that is large enough
to adequately perform the proposal update and provide consistent
estimates of the variables of interest. The new technique is  introduced in Section \ref{Algorithms} below.

\section{Algorithms}
\label{Algorithms}

In this section we describe the proposed algorithm, which is termed nonlinear PMC (NPMC). We adopt a very simple proposal update scheme, where the importance functions are multivariate Gaussian pdf's with moments matched to our latest approximation of the posterior distribution. The key feature is the application of a nonlinear transformation of the importance weights. Besides the basic version of the algorithm, we propose an adaptive version where this transformation is only applied when the value of the effective sample size is below a certain threshold. Finally, we explore different forms of the weight transformation.

\subsection{Nonlinear PMC}

Assume, in the sequel, that the target pdf is the posterior density, i.e., $\pi(\boldsymbol{\theta}) \propto p(\boldsymbol{\theta}|\textbf{y})p(\boldsymbol{\theta})$. For simplicity we select the importance functions in the PMC
scheme as multivariate normal (MVN) densities. In the $\ell$-th iteration
\begin{equation*}
q_\ell(\boldsymbol{\theta}) = \mathcal{N} \left(
\boldsymbol{\theta}; \boldsymbol{\mu}_\ell, \boldsymbol{\Sigma}_\ell
\right), \quad \ell = 1, \dots, L,
\end{equation*} 
where $\boldsymbol{\mu_\ell}$ is the mean vector and $\boldsymbol{\Sigma}_\ell$ is a positive definite covariance marix. The parameters of the Gaussian proposal are chosen to match the moments of the distribution described by the discrete measure
ontained after the $(\ell-1)$-th iteration of the PMC algorithm. In particular, we compute the mean and covariance as 
\begin{eqnarray}
\boldsymbol{\mu}_{\ell} &=& \frac{1}{M} \sum_{i=1}^M
\tilde{\boldsymbol{\theta}}_{\ell-1}^{(i)} \quad \mbox{and} \label{eqCompMean}\\
\boldsymbol{\Sigma}_{\ell} &=& \frac{1}{M} \sum_{i=1}^M \left(
\tilde{\boldsymbol{\theta}}_{\ell-1}^{(i)} - \boldsymbol{\mu}_{\ell}
\right)\left( \tilde{\boldsymbol{\theta}}_{\ell-1}^{(i)} -
\boldsymbol{\mu}_{\ell} \right)^\top. \label{eqCompCov}
\end{eqnarray}
Let us remark that this particular proposal update scheme is not a constraint of the algorithm. It is actually independent of the weight update and resampling steps and can be designed as freely as in the standard PMC methodology.

The key modification of the algorithm is the computation of the importance weights. Given a variate $\boldsymbol{\theta}^{(i)}_\ell$, its asociated weight is computed as $\bar w_\ell^{(i)} \propto \varphi_\ell^M( w_\ell^{(i)*}$, where $w_\ell^{(i)*}=p(\boldsymbol{\theta}|\textbf{y})p(\boldsymbol{\theta}) / q_\ell({\boldsymbol{\theta}}^{(i)}_\ell$ is the standard unnormalized importance weight and $\varphi_\ell^M : (0,\infty) \rightarrow (0,\infty)$ is a nonlinear positive function that may depend both on the iteration index $\ell$ and the number of samples $M$. This nonlinearity should be chosen so as to reduce the variation of the resulting normalized weights, $\bar w_\ell^{(i)} = w_\ell^{(i)*} / \sum_{j=1}^M w_\ell^{(j)*}$. Intuitively, it should preserve the ordering of the samples (those with larger standard weights should also have the largest transformed weights) while reducing the difference $\max_i \bar w_\ell^{(i)} - \min_i \bar w_\ell^{(i)}$ or some other measure of weight variation. The proposed generic algorithm is outlined in Table \ref{tNPMCalgorithm}.

\begin{table}
\caption{Nonlinear PMC algorithm}

\vspace{0.2cm}
\underline{\textbf{Iteration ($\ell = 0, \ldots, L$):}}

\begin{enumerate}
\item Draw a collection of $M$ samples
$\Theta_\ell^M = \left\{ \boldsymbol{\theta}_\ell^{(i)}
\right\}_{i=1}^M$ from the proposal density
$q_\ell(\boldsymbol{\theta})$.

\begin{itemize}
\item At iteration $\ell=0$, the proposal density is the prior pdf $q_0
(\boldsymbol{\theta}) = p(\boldsymbol{\theta})$.

\item At iterations $\ell=1,\ldots,L$ the proposal is a MVN pdf $q_\ell(\boldsymbol{\theta}) =
\mathcal{N} (\boldsymbol{\theta}; \boldsymbol{\mu}_\ell,
\boldsymbol{\Sigma}_\ell)$, where the mean and covariance are computed according to Eqs. (\ref{eqCompMean}) and (\ref{eqCompCov}).
\end{itemize}

\item Compute the unnormalized PMC weights
\begin{equation*}
w_\ell^{(i)*} = \frac{p \left( \boldsymbol{\theta}_\ell^{(i)}
\left|\right. \textbf{y} \right)}{q_\ell \left(
\boldsymbol{\theta}_\ell^{(i)} \right)} \propto \frac{ p \left(
\textbf{y} \left|\right. \boldsymbol{\theta}_\ell^{(i)} \right) p
\left( \boldsymbol{\theta}_\ell^{(i)} \right) }{q_\ell \left(
\boldsymbol{\theta}_\ell^{(i)} \right)}, \quad i=1,\ldots, M.
\end{equation*}

\item Perform non-linear transformations $\varphi_\ell^M$ on the weights in order to smooth their
variations, that is,
\begin{equation*}
\bar{w}_\ell^{(i)*} = \varphi_\ell^M \left( w_\ell^{(i)*} \right), \quad i=1,\ldots, M.
\end{equation*}

\item Normalize the modified importance weights,
\begin{equation*}
\bar{w}_\ell^{(i)} = \frac{\bar{w}_\ell^{(i)*}}{\sum_{j=1}^M
\bar{w}_\ell^{(j)*}}, \quad i=1, \ldots, M.
\end{equation*}

\item Resampling: for $i = 1, \ldots, M$, let $\tilde{\boldsymbol{\theta}}_\ell^{(i)} = \boldsymbol{\theta}_\ell^{(j)}$
with probability $w_\ell^{(j)}$, $j=1,\ldots,M$. We obtain $\tilde \Theta_\ell^M$.

\end{enumerate}
\vspace{0.2cm}
\label{tNPMCalgorithm}
\end{table}

Step 5 of the NPMC method involves multinomial resampling, which
consists in sampling with replacement from the set $\{
\boldsymbol{\theta}_\ell^{(i)}\}_{i=1}^M$ with probabilities
equal to the associated weights $w_\ell^{(i)}$, to obtain an
unweighted set $\{ \tilde{\boldsymbol{\theta}}_\ell^{(i)}\}_{i=1}^M$. This is obviously not the only choice of resampling algorithm and we use it only for the sake of simplicity. See, e.g., \cite{Douc07} for an overview of resampling
techniques.

At each iteration $\ell = 0, \ldots, L$, we obtain random discrete approximations of the posterior distribution with density $\pi(\boldsymbol{\theta})$. Indeed, the discrete measures
\begin{eqnarray}
\bar{\pi}_\ell^M(d\boldsymbol{\theta}) &=& \sum_{i=1}^M
\bar{w}_\ell^{(i)} \delta_{\boldsymbol{\theta}_\ell^{(i)}} (d\boldsymbol{\theta}), \quad \mbox{and} \nonumber\\
\tilde{\pi}_\ell^M(d\boldsymbol{\theta}) &=& \frac{1}{M} \sum_{i=1}^M
\delta_{\boldsymbol{\theta}_\ell^{(i)}} (d\boldsymbol{\theta})
\end{eqnarray}
are, both of them, approximations of $\pi(\boldsymbol{\theta})d\boldsymbol{\theta}$. In particular, if $f:\mathbb{R}^K\rightarrow\mathbb{R}$ is an arbitrary integrable function of $\boldsymbol{\theta}$, then the integral $(f,\pi)=\int f(\boldsymbol{\theta})\pi(\boldsymbol{\theta})d\boldsymbol{\theta}$ can be approximated as either 
\begin{eqnarray}
(f,\pi) \approx (f,\bar \pi_\ell^M) = \sum_{i=1}^M f({\boldsymbol{\theta}}^{(i)}_\ell) \bar w_\ell^{(i)} \quad \mbox{or} \nonumber\\
(f,\pi) \approx (f,\tilde \pi_\ell^M) = \frac{1}{M} \sum_{i=1}^M f({\boldsymbol{\theta}}^{(i)}_\ell). \nonumber
\end{eqnarray}

Note that the estimator $(f,\tilde \pi_\ell^M)$ involves one extra Monte Carlo step (because of resampling) and, hence, it can be shown to have more variance than $(f,\bar \pi_\ell^M)$ \cite{Douc07}. Therefore, we assume in the sequel that estimates are computed by way of the measure $\bar \pi_\ell^M$ unless explicitly stated otherwise. 

Note as well that, since $\pi(\boldsymbol{\theta}) \propto p(\boldsymbol{\theta}|\textbf{y})p(\boldsymbol{\theta})$, any expectation with respect to the posterior distribution is actually an integral with respect to the measure $\pi(\boldsymbol{\theta})d\boldsymbol{\theta}$, i.e., $E_{p(\boldsymbol{\theta}|\textbf{y}}[ f(\boldsymbol{\theta}) ]  = (f,\pi)$, and, therefore, it can be approximated using $\bar \pi_\ell^M$, namely, $E_{p(\boldsymbol{\theta}|\textbf{y}}[ f(\boldsymbol{\theta}) ]  = (f,\bar \pi_\ell^M)$.

\subsection{Modified NPMC}

The nonlinear transformation $\varphi_\ell^M$ is mainly useful at the first iterations of the PMC scheme, when the proposal
density does not fit the target density and the standard importance weights may display high variability.
However, in some applications it may be possible to remove the nonlinear transformation after a few iterations, when the proposal is closer to the target.

Table \ref{tNPMCModified} displays a modification of the NPMC algorithm which consists in
applying the nonlinear transformation only if the ESS is below a specified threshold. The modified algorithm only
differs in steps 3 and 4 from the generic NPMC procedure, and they are outlined in the table.

\begin{table}
\caption{Modified NPMC algorithm}
\vspace{0.2cm}
Steps 3 and 4 of the NPMC algorithm are replaced by the following computations:
\begin{itemize}
\item[3.] Compute the normalized importance weights $w_\ell^{(i)}$ and the corresponding ESS
$M_\ell^{eff}$
\begin{equation*}
M_\ell^{eff} = \frac{1}{\sum_{i=1}^M \left( w_\ell^{(i)} \right)^2}.
\end{equation*}

\item[4.] If $M_\ell^{eff} < M_{min}^{eff}$, smooth and normalize the weights
\begin{equation*}
\bar{w}_\ell^{(i)*} = \varphi_\ell^M \left( w_\ell^{(i)*} \right),
\; \bar{w}_\ell^{(i)} = \frac{\bar{w}_\ell^{(i)*}}{\sum_{j=1}^M
\bar{w}_\ell^{(j)*}}, \; i=1, \ldots, M.
\end{equation*}
Otherwise, set $\bar{w}_\ell^{(i)} = w_\ell^{(i)}$.
\end{itemize}
\vspace{0.2cm}
\label{tNPMCModified}
\end{table}

\subsection{Selecting the transformation of the importance weights}

Several possibilities exist for the choice of the nonlinearity $\varphi_\ell^M$. In this section we describe and intuitively justify two specific functions based on the ``tempering'' and the ``clipping'', respectively, of the standard importance weights.

\subsubsection{Tempering}

In this case, the nonlinear transformation $\varphi_\ell^M$ consists
in an exponentiation of the standard weights. In particular, the expression of the transformed weights 
\begin{equation*}
\bar{w}_\ell^{(i)*} = \varphi_\ell^M \left( w_\ell^{(i)*} \right) =
\left( w_\ell^{(i)*}
\right)^{\gamma_\ell}, \quad i=1,\ldots, M,
\end{equation*}
where $0<\gamma_\ell\le 1$. The sequence $\gamma_\ell$, $\ell=1,\ldots,L$, has to be
adapted along the iterations, taking low values at the first steps and getting closer to 1 (an, eventually, $\gamma_\ell=1$) as the algorithm converges.

A straightforward a priori choice for the $\gamma_\ell$ sequence is a
polinomial function of $\ell$, e.g.,
$\gamma_\ell \propto \ell^m$, or a sigmoid function $\gamma_\ell =
\frac{1}{1-e^{-\ell}}$. However, this choices do not guarantee that the resulting ESS be large enough. While in simple examples this procedure provides a remarkable
reduction of the weight variations and an increase of the ESS,
in more complex problems it is not enough to guarantee a
stable and consistent convergence, as will be shown numerically in Sections
\ref{GMM_example} and \ref{SKM_example}. In such cases, the sequence
$\gamma_\ell$ may be computed adaptively, according to the values taken by the standard importance
weights $w_\ell^{(i)}$, $i=1, ..., M$.

This methodology can be interpreted as a generalization of the \textit{simulated
tempering} process applied to the target density, which has been
widely studied in the MCMC literature, \cite{Robert04}. However, to
the best of our knowledge, it has not been applied in iterative IS
approaches. The technique can also related to the one proposed in
\cite{Koblents11}, which introduces a sequence of models
\begin{equation*}
\pi_\ell(\boldsymbol{\theta}) =
p_\ell(\boldsymbol{\theta}|\textbf{y}) \propto p_\ell(\textbf{y} |
\boldsymbol{\theta}) p(\boldsymbol{\theta}), \quad \ell=1,\ldots,L
\end{equation*}
that converges to the true posterior pdf
$p(\boldsymbol{\theta}|\textbf{y})$ along the iterations. These
models are constructed in such a way that they are simpler for Monte
Carlo approximation, since they have a broader likelihood and
partially avoid degeneracy.

\subsubsection{Clipping}
\label{clipping}

In problems where the degeneracy of the weights is severe it is
often difficult to apply a tempering procedure to soften the weights
and to obtain a reasonable ESS. The softening factor must take
extremely low values at the first iterations and a prior selection
of the sequence $\gamma_\ell$ is not straightforward.

In this work we propose a simple technique that consists in clipping the standard weights that are above a certain threshold. As a consequence, a sufficient number of ``flat'' (modified) weights is guaranteed in the regions of the space of theta where the standard weights were larger. The ESS becomes correspondingly larger as well.

Specifically, the modified weights at iteration $\ell$ are computed
from the original importance weights as
\begin{equation*}
\bar{w}_\ell^{(i)*} = \min \left( w_\ell^{(i)*} ,
\mathcal{T}_\ell^M \right), \quad i=1,\ldots,M,
\end{equation*}
where the threshold $\mathcal{T}_\ell^M$ is selected to guarantee that the
number of samples $\boldsymbol{\theta}_\ell^{(i)}$ that satisfy
$w_\ell^{(i)*} \geq \mathcal{T}_\ell^M$ is equal to $M_T<M$. The
parameter $M_T$ must be selected to represent adequately the
multidimensional target posterior pdf $\pi(\boldsymbol{\theta}) = p
\left( \boldsymbol{\theta} | \textbf{y} \right)$.

An alternative approach to the hard clipping technique described
above, is to apply a sigmoid transformation to the standard weights, resulting in a soft
clipping, namely
\begin{equation*}
\bar{w}_\ell^{(i)^*} = \frac{2 \beta_\ell}{1 + \exp \left( - \frac{2
w_\ell^{(i)*}} {\beta_\ell} \right) } - \beta_\ell, \quad i=1,\ldots,M,
\end{equation*}
where $\beta_\ell > 0$ should increase along the iterations in order to progressively reduce the
nonlinear distortion of the standard weights.

\section{Convergence of nonlinear importance sampling}
\label{Analysis}

The convergence of the original PMC scheme is easily justified by the convergence of the standard IS method. Indeed, it can be proved  that the discrete measure $\pi_\ell^M(d\boldsymbol{\theta}) = \sum_{i=1}^M w_\ell^{(i)} \delta_{{\boldsymbol{\theta}}_\ell^{(i)}}(d\boldsymbol{\theta})$ converges to $\pi(\boldsymbol{\theta})d\boldsymbol{\theta}$ under mild assumptions, meaning that
\begin{equation}
\lim_{M \rightarrow \infty} | (f,\pi_\ell^M) - (f,\pi) | = 0
\label{eqConvStdIS1}
\end{equation}
almost surely (a.s.) for every $\ell \in \{1, ..., L \}$ and any $f \in B(\mathbb{R}^K)$. 

In this section we provide a result similar to Eq. (\ref{eqConvStdIS1}) for the discrete measure $\bar \pi_\ell^M$ generated by the NPMC algorithm using a class of clipping transformations. The analysis, therefore, is concerned with the asymptotic performance of the approximation as the number of amples $M$ grows, and not with the iteration of the algorithm, i.e., not with the convergence as $\ell$ increases. Hence, we shall drop the latter subscript for convenience in the sequel.

\subsection{Notation and basic assumptions}

Let $\pi$ be the pdf associated to the target probability distribution to be approximated and let $q$ be the importance function used to propose samples in an IS scheme (not necessarily normalized) and let $h(\boldsymbol{\theta}) = a \pi(\boldsymbol{\theta})$ be a function proportional to $\pi$, with the proportionality constant $a$ independent of $\boldsymbol{\theta}$. The samples drawn from the distribution associated to $q$ are denoted $\boldsymbol{\theta}^{(i)}$, $i=1, ..., M$, while their associated unnormalized importance weights are
\begin{equation*}
w^{(i)*} = \frac{
	h( {\boldsymbol{\theta}}^{(i)} )
}{
	q( {\boldsymbol{\theta}}^{(i)} )
}, \quad i=1, ..., M.
\end{equation*}

Let us define the weight function $g=h/q$, i.e., $g(\boldsymbol{\theta}) = h(\boldsymbol{\theta})/q(\boldsymbol{\theta})$ and, in particular, $g( {\boldsymbol{\theta}}^{(i)}) = w^{(i)*}$. The support of $g$ is the same as the support of $q$, denoted $\mathbb{S} \subseteq \mathbb{R}^K$. If we assume that both $q(\boldsymbol{\theta})>0$ and $\pi(\boldsymbol{\theta})\le0$ for any $\boldsymbol{\theta} \in \mathbb{S}$, then $g(\boldsymbol{\theta})\ge 0$ for every $\boldsymbol{\theta} \in \mathbb{S}$ as well. Also, trivially, $\pi \propto gq$, with the proportionality constant independent of $\boldsymbol{\theta}$. These assumptions are standard for classical IS.

The approximation of the target probability measure generated by the standard IS method is 
\begin{equation*}
\pi^M(d\boldsymbol{\theta}) = \sum_{i=1}^M w^{(i)} \delta_{{\boldsymbol{\theta}}^{(i)}}(d\boldsymbol{\theta}),
\end{equation*}
where 
\begin{equation*}
w^{(i)} = \frac{
	g({\boldsymbol{\theta}}^{(i)})
}{
	\sum_{j=1}^M g({\boldsymbol{\theta}}^{(j)})
}
\end{equation*}
is the $i$-th normalized standard weight.

The nonlinear transformation $\varphi^M$ of the weights is assumed to be of a clipping class. In particular, let $i_1, i_2, ..., i_M$ be a permutation of the indices in $\{1, ..., M \}$ such that $w^{(i_1)*} \le w^{(i_2)*} \le \ldots \le w^{(i_M)*}$ and choose $M_T < M$ and some constant $C < \infty$. Then, the transformed weights satisfy
\begin{equation}
\left.
	\begin{array}{ll}
	\bar w^{(i_k)*} &= \varphi^M(w^{(i_k)*}) \le C, \quad \mbox{for } k=1, \ldots, M_T, \quad \mbox{and}\\
	\bar w^{(i_k)*} &= \varphi^M(w^{(i_k)*}) = w^{(i_k)*}, \quad \mbox{for } k=M_T+1, \ldots, M\\
	\end{array}
\right\} \label{eqConditionPhi}
\end{equation}  
The condition above is can be satisfied for various forms of hard and soft clipping when the weight function $g$ is upper bounded. In particular, if $g \in B(\mathbb{R}^K)$ then, trivially, $\bar w^{(i)*} \le \| g \|_\infty$ for either the hard or the soft clipping procedures described in Section \ref{clipping} (hence, $C \le \| g \|_\infty$).

The approximation of the target probability measure generated by the nonlinear IS method is
\begin{equation*}
\bar \pi^M(d\boldsymbol{\theta}) = \sum_{i=1}^M \bar w^{(i)} \delta_{{\boldsymbol{\theta}}^{(i)}}(d\boldsymbol{\theta}),
\end{equation*}
where 
\begin{equation*}
\bar w^{(i)} = \frac{
	\varphi^M( g({\boldsymbol{\theta}}^{(i)}) )
}{
	\sum_{j=1}^M \varphi( g({\boldsymbol{\theta}}^{(j)}) )
}
\end{equation*}
is the $i$-th normalized transformed weight. Additionally, we introduce a non normalized approximation of $\pi(\boldsymbol{\theta})d\boldsymbol{\theta}$ of the form
\begin{equation}
\check \pi^M(d\boldsymbol{\theta}) = \sum_{i=1}^M \check w^{(i)} \delta_{{\boldsymbol{\theta}}^{(i)}}(d\boldsymbol{\theta}),
\label{eqDefBridge}
\end{equation}
where
\begin{equation*}
\check w^{(i)} = \frac{
	\varphi^M( g({\boldsymbol{\theta}}^{(i)}) )
}{
	\sum_{j=1}^M g({\boldsymbol{\theta}}^{(j)})
}, \quad i=1, ..., M,
\end{equation*}
are non normalized transformed weights that will be referred to as ``bridge weights'' in the sequel.

\subsection{Asymptotic convergence}

We aim at proving that $\lim_{M\rightarrow \infty} | (f,\bar \pi^M) - (f,\pi) | = 0$ a.s. for any $f \in B(\mathbb{R}^K)$. To obtain such a result, we split the problem into simpler questions by applying the triangle inequality
\begin{equation}
| (f,\bar \pi^M) - (f,\pi) | \le | (f,\bar \pi^M) - (f,\pi^M) | + | (f,\pi^M) - (f,\pi) |.
\label{eqTriangle-1}
\end{equation}
The second term on the right hand side of (\ref{eqTriangle-1}) is handled easily using standard IS theory. For the first term, we have to prove that the discrete measure generated by the {\em nonlinear} IS method ($\bar \pi^M$) converges to the discrete measure generated by the standard IS method ($\pi^M$). This can be done by resorting to another triangle inequality,
\begin{equation}
| (f,\bar \pi^M) - (f,\pi^M) | \le | (f,\bar \pi^M) - (f,\check \pi^M) | + | (f,\check \pi^M) - (f,\pi^M) |,
\label{eqTriangle-2}
\end{equation}
that reveals the role of the bridge measure defined in (\ref{eqDefBridge}).

The following lemma establishes the asymptotic convergence of the term $| (f,\bar \pi^M) - (f,\check \pi^M) |$ in inequality (\ref{eqTriangle-2}).

\begin{lemma} \label{L1}
Assume that $\lim_{M \rightarrow \infty} \frac{M_T}{M} = 0$, $g \in B(\mathbb{R}^K)$, $g > 0$ and the transformation function $\varphi^M$ satisfies (\ref{eqConditionPhi}). Then, for every $f \in B(\mathbb{R}^K)$ and sufficiently large $M$, there exist positive constants $c_1, c_1'$ independent of $M$ and $M_T$ such that
\begin{equation*}
\mathbb{P}\left\{ 
		\left| 
			( f,\bar \pi^M ) - (f,\check \pi^M)
		\right| 
	\leq c_1 \frac{M_T}{M} 
\right\} 
\geq
1 - \exp{ \left( -c_1'M \right)}.
\end{equation*}
\end{lemma}
\noindent {\bf Proof}: See Appendix \ref{App_1}.

Next, we establish the convergence of the bridge measure $\check \pi^M$ toward $\pi^M$.

\begin{lemma} \label{L2}
Assume that $g \in B(\mathbb{R}^K)$, $g>0$ and the transformation function $\varphi^M$ satisfies (\ref{eqConditionPhi}). Then, for every $f \in B(\mathbb{R}^K)$ there exist positive constants $c_2, c_2'$ independent of $M$ and $M_T$ such that
\begin{equation*}
\mathbb{P}\left\{ 
		\left| 
			( f,\check \pi^M ) - (f,\pi^M)
		\right| 
	\leq c_2 \frac{M_T}{M} 
\right\} 
\geq
1 - \exp{ \left( -c_2'M \right)}.
\end{equation*}
\end{lemma}
\noindent{\bf Proof}: See Appendix \ref{App_2}.

The combination of Lemmas \ref{L1} and \ref{L2}, together with the triangle inequality (\ref{eqTriangle-2}), yields the convergence of the error $| (f,\bar \pi^M) - (f,\pi^M) |$.

\begin{lemma} \label{L3}
Assume that $\lim_{M\rightarrow \infty} \frac{M_T}{M}=0$, $g \in B(\mathbb{R}^K)$, $g>0$ and the transformation function $\varphi^M$ satisfies (\ref{eqConditionPhi}). Then, for every $f \in B(\mathbb{R}^K)$, and sufficiently large $M$, there exist positive constants $c, c'$ independent of $M$ and $M_T$ such that
\begin{equation*}
\mathbb{P}\left\{ 
		\left| 
			( f,\bar \pi^M ) - (f, \pi^M)
		\right| 
	\leq c \frac{M_T}{M} 
\right\} 
\geq
1 - 2\exp{ \left( -c'M \right)}.
\end{equation*}
In particular,
\begin{equation*}
\lim_{M\rightarrow\infty} | (f,\bar \pi^M) - (f,\pi^M) | = 0 \quad \mbox{a.s.}
\end{equation*}
\end{lemma}
\noindent {\bf Proof}: See Appendix \ref{App_3}.

Finally, Lemma \ref{L3} can be combined with inequality (\ref{eqTriangle-1}) to yield the desired result, stated below.

\begin{theorem} \label{Th1}
Assume that $g \in B(\mathbb{R}^K)$ and the transformation function $\varphi^M$ satisfies (\ref{eqConditionPhi}) with $C \le \| g \|_\infty$ independent of $M$. If $\lim_{M\rightarrow\infty} \frac{M_T}{M}=0$ then, for every $f \in B(\mathbb{R}^K)$,
\begin{equation*}
\lim_{M\rightarrow\infty} | (f,\bar \pi^M) - (f,\pi) | = 0 \quad \mbox{a.s.}
\end{equation*}
\end{theorem}
\noindent {\bf Proof}: It is classical result that 
\begin{equation}
\lim_{M\rightarrow \infty} | (f,\pi^M) - (f,\pi) | = 0 \quad \mbox{a.s.}
\label{eqWellKnown}
\end{equation}
Combining (\ref{eqWellKnown}) with the second part of Lemma \ref{L3} and the triangle inequality in (\ref{eqTriangle-1}) yields the desired result. \qed

\section{Example 1: a Gaussian mixture model}
\label{GMM_example}

In this section we provide numerical results that illustrate the degeneracy
problem and the performance of the proposed NPMC scheme applied to
the Gaussian mixture model example of \cite{Cappe04}.

\subsection{Problem setup}
\label{GMM_probsetup}

We consider the Gaussian mixture model given by
\begin{equation}
\label{GMM} p(y | \boldsymbol{\theta}) = \rho \mathcal{N} \left( y;
\theta_1, \sigma^2 \right) + (1-\rho) \mathcal{N} \left( y;
\theta_2, \sigma^2 \right)
\end{equation}
where the variable of interest $\boldsymbol{\theta} = [\theta_1$,
$\theta_2]^\top$ has dimension $K=2$, and contains the means of the
mixture components. The true values of the unknowns are set to
$\boldsymbol{\theta} = [0,2]^\top$. The mixture coefficient and the
variance of the components are assumed to be known and set to $\rho
= 0.2$ and $\sigma^2=1$.

We assume a prior pdf $p(\boldsymbol{\theta}) = p(\theta_1)
p(\theta_2)$ composed of equal independent components for each
unknown, given by $p(\theta_k) = \mathcal{N} \left( \theta_k ; \nu,
\sigma^2/\lambda \right)$, for $k=1,2$. The hyperparameters are also
assumed to be known and set to $\nu = 1$ and $\lambda = 0.1$.

We consider that a set $\textbf{y}$ of $N$ iid scalar observations drawn
from the mixture model in equation (\ref{GMM}) is available. We thus
aim at approximating the target density 
$\pi(\boldsymbol{\theta})=p(\boldsymbol{\theta} | \textbf{y}) \propto p(\textbf{y} |\boldsymbol{\theta}) p(\boldsymbol{\theta})$, i.e., the posterior of
$\boldsymbol{\theta}$ given the set of observations $\textbf{y}$.

\subsection{Degeneracy of the Importance weights}

In this section we analyze the degeneracy problem, addressed in Section \ref{Degeneracy}, in a simple
and low dimensional IS example.

In order to illustrate the effects of degeneracy, in this
section we focus on the standard importance sampling procedure. We
consider a set of $M$ samples $\Theta^M = \{
\boldsymbol{\theta}^{(i)} \}_{i=1}^M$ drawn form the prior pdf
$p(\boldsymbol{\theta})$. Thus, the normalized importance weights
are computed from the likelihood function as
\begin{equation*}
w^{(i)} \propto p \left( \textbf{y} | \boldsymbol{\theta}^{(i)}
\right) = \prod_{n=1}^N p \left( y_n | \boldsymbol{\theta}^{(i)}
\right) =
\end{equation*}
\begin{equation*}
 = \prod_{n=1}^N \rho \mathcal{N} \left( y_n;
\theta_1^{(i)}, \sigma^2 \right) + (1-\rho) \mathcal{N} \left( y_n;
\theta_2^{(i)}, \sigma^2 \right)
\end{equation*}

\begin{figure*}[t]
\centering 
\includegraphics[width=0.49\textwidth]{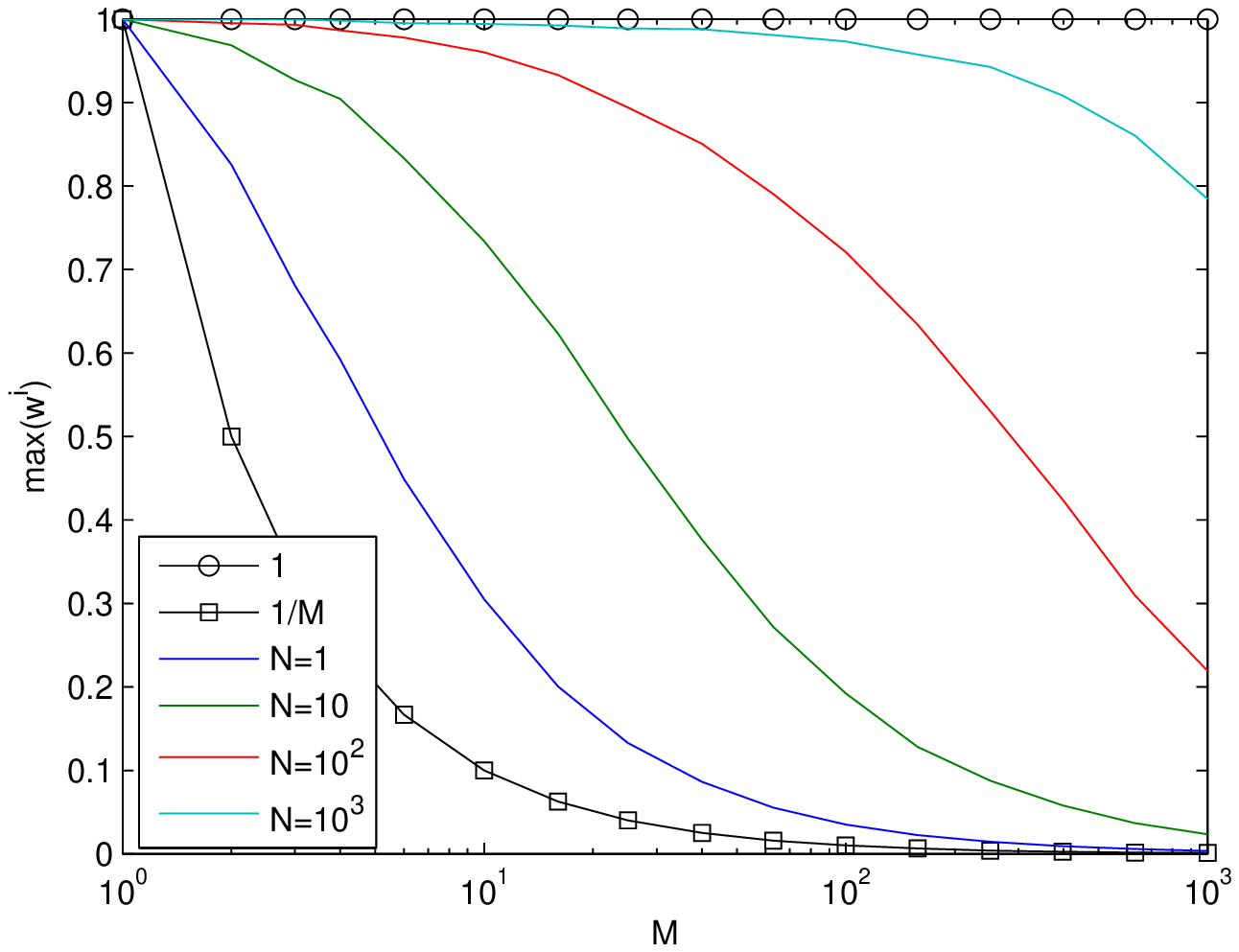}
\includegraphics[width=0.49\textwidth]{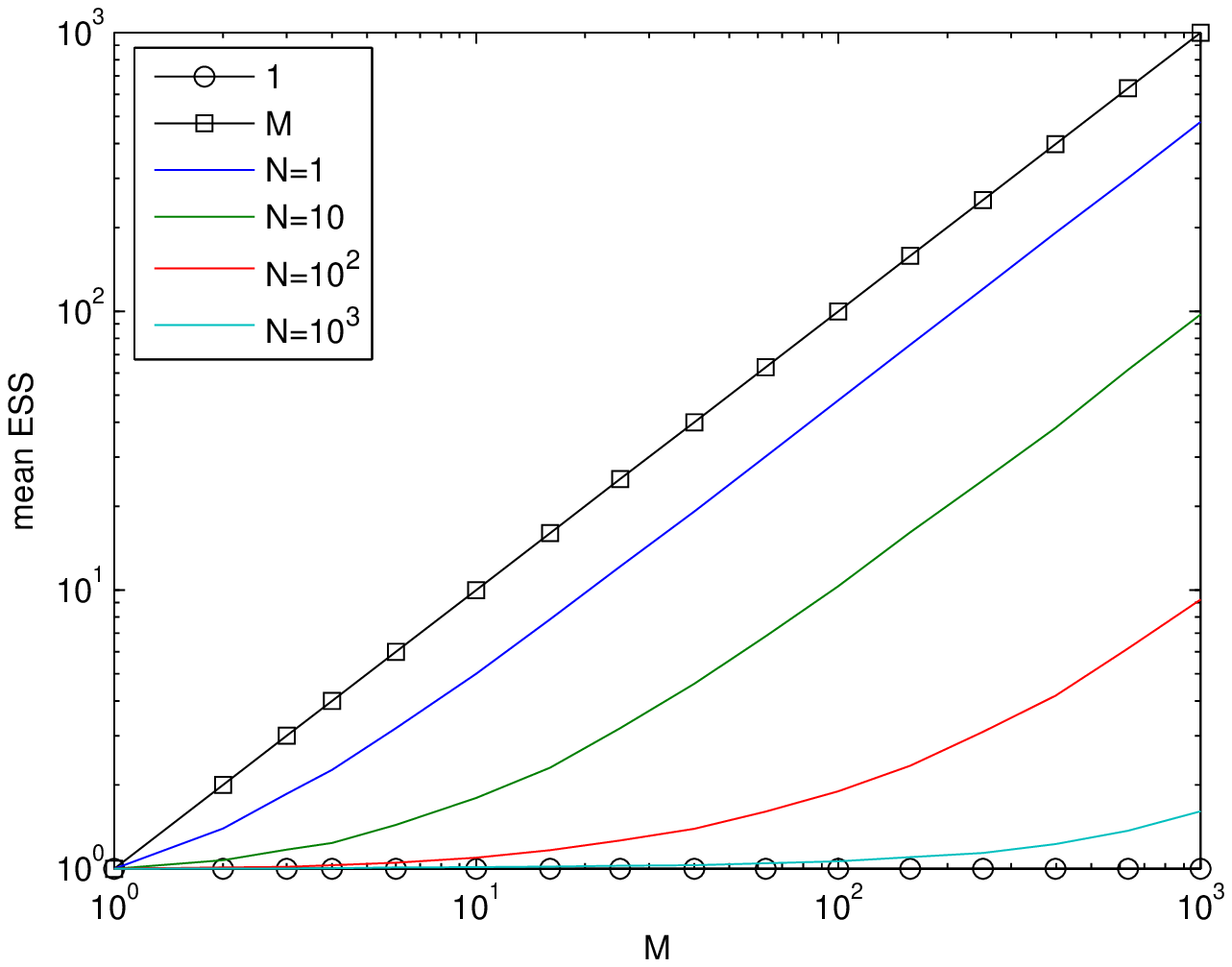}
\caption{Evolution of the maximum importance weight
(\emph{left}) and the ESS (\emph{right}) vs. the number of
observations, $N$, and the number of samples, $M$. The curves corresponding to the
best possible case(uniform weights and $M$ effective samples) are
depicted with squares. The curves representing extreme degeneracy ($\max_i w^{(i)} = 1$ and ESS$=1$) are plotted
with circles. The results displayed are averaged over $P=10^3$ independent
simulation runs.} \label{ESS_MPMC}
\end{figure*}

For this model, we have analyzed the behavior of the maximum normalized importance
weight $\max w^{(i)}$ and the effective sample size $M^{eff}$, when the number of observations $N$
increases. We consider a number of observations $N$ varying from 1
to $10^3$, and a number of samples $M$ also varying from 1 to
$10^3$. For each pair of $N$ and $M$ we have performed $P=10^3$
simulation runs, generating an independent vector of
observations $\textbf{y}$ and an independent set of samples $\Theta^M$
in each run.

In Figure \ref{ESS_MPMC} (left) the maximum importance weight
averaged over $P$ simulation runs is represented as a function of
the number of samples $M$ and the number of observations $N$. The
curves representing the extreme cases $\max_i w^{(i)} = 1/M$ (uniform weights) and $\max_i
w^{(i)} = 1$ (degeneracy) are also plotted on the graph. It can be
observed that, as the number of observations, $N$, increases, and
considering a fixed number of samples $M$, the maximum importance
weight moves apart from the optimal value $1/M$ and
approaches degeneracy, where $\max_i w^{(i)}=1$. This
indicates that an increase in the number of observations causes an
increase in the fluctuations of the importance weights.

On the other hand, in Figure \ref{ESS_MPMC} (right) the average
ESS is also represented for several values of $M$ and $N$. The cases when
the ESS is maximum, $M^{eff}=M$, and minimum, $M^{eff} = 1$, are
plotted on the graph for reference. It can be observed that, as the number
of observations $N$ increases, the ESS is smaller for the same value of $M$. 
Actually, while the ESS always grows with $M$, its slope becomes very small
for large values of $N$. For example, with $N=1000$ observations and 
$M=1000$ samples, we only obtain 1.5 effective samples
on average. 

\begin{figure}[t]
\centering 
\includegraphics[width=0.49\textwidth]{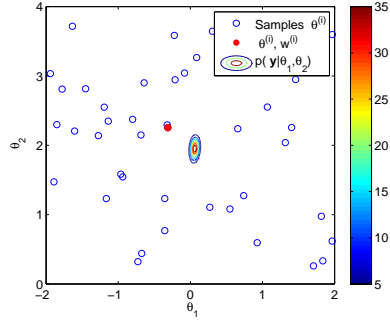}
\caption{Subset of $M=200$ samples $\boldsymbol{\theta}^{(i)}$ drawn
form the prior $p(\boldsymbol{\theta})$ (blue circles) and the
associated standard importance weights (red circles with size proportional to
the weight $w^{(i)}$). The likelihood function
$p(\textbf{y}|\boldsymbol{\theta})$ is depicted with contour lines.}
\label{GMM_samples}
\end{figure}

In Figure \ref{GMM_samples} an example of degeneracy in this simple
setup is shown. A set of $M=200$ samples has been drawn from the
prior pdf and standard importance weights have been computed
from the likelihood. A subset of these samples is depicted in figure
\ref{GMM_samples} together with the associated weights. The
likelihood function evaluated in the same region is depicted with
contour lines and it presents high variations due to the high number
of observations. It can be observed that only a small fraction of the
sample set is close to the region where the likelihood is significant. As a
result, one sample has a weight close to 1 and the rest of them have
negligible weights.

\subsection{Comparison of algorithms}

In this section we compare, by way of computer simulations, the performance of the two proposed  schemes (NPMC with tempering and with clipping, see Section \ref{Algorithms}) and the standard PMC algorithm when applied to the Gaussian mixture model. For convenience, we reproduce in Table \ref{tReproducePMC} the original PMC scheme proposed in \cite{Cappe04}.



\begin{table}
\caption{Standard PMC algorithm of \cite{Cappe04}.}  \vspace{0.1cm}
\underline{\textbf{Initialization ($\ell=0$):}}
\begin{enumerate}
\item Consider a set of $p$ scales (variances) $v_j$ and an initial number
$r_j = m$ of samples per scale, $j=1, \ldots,p$.
\item For $i=1,\ldots,M=pm$, draw $\left\{ \boldsymbol{\theta}_0^{(i)} \right\}$ from $q_0(\boldsymbol{\theta}) =
p(\boldsymbol{\theta})$. 
\end{enumerate}

\vspace{0.1cm} \underline{\textbf{Iteration ($\ell = 1,\ldots,L$):}}
\begin{enumerate}
\item For $j=1,\ldots,p$
\begin{itemize}
\item generate a sample $\left\{ \boldsymbol{\theta}_\ell^{(i)} \right\}$ of size $r_j$
from
\begin{equation*}
q_\ell \left( \boldsymbol{\theta} \right) = \mathcal{N}
\left(\boldsymbol{\theta}_\ell^{(i)};
\boldsymbol{\theta}_{\ell-1}^{(i)}, v_j \textbf{I}_K \right), \quad
i = 1,\ldots, M,
\end{equation*}

\item compute the normalized weights
\begin{equation*}
w_\ell^{(i)} \propto \frac{p \left( \textbf{y} |
\boldsymbol{\theta}_\ell^{(i)} \right) p \left(
\boldsymbol{\theta}_\ell^{(i)} \right)}{q_\ell \left(
\boldsymbol{\theta}_\ell^{(i)} \right)}.
\end{equation*}
\end{itemize}

\item Resample with replacement the set $\left\{ \boldsymbol{\theta}_\ell^{(i)} \right\}_{i=1}^M$
according to the weights $w_\ell^{(i)}$ to obtain
$\left\{\tilde{\boldsymbol{\theta}}_\ell^{(i)}\right\}_{i=1}^M$.

\item For $j=1,\ldots,p$ update $r_j$ as the number of elements
generated with variance $v_j$ which have been resampled.
\end{enumerate}
\vspace{+0.2cm}
\label{tReproducePMC}
\end{table}

In order to asses the merit of an unweighted sample set $\{
\tilde{\theta}_k^{(i)} \}_{i=1}^M$ for the estimation of a scalar variable of interest 
$\theta_k$, we evaluate the mean square error (MSE)
\begin{equation*}
MSE_k = \frac{1}{M} \sum_{i=1}^M \left( \tilde{\theta}_k^{(i)} -
\theta_k \right)^2,
\end{equation*}
where $M$ is the sample size and $k \in \{ 1, ..., K \}$.

The  parameters common to all the algorithms have been set to $M=200$
samples per iteration and $L=20$ iterations. We have performed
$P=10^3$ independent simulation runs, with an independent random vector of
observations $\textbf{y}$ at each run. The standard PMC algorithm and the proposed
NPMC technique, both with tempering and clipping (labeled {\em NPMC temp} and {\em NPMC clip}, respectively, in the figures)
have been run with the same observation vector
$\textbf{y}$ and the results have been compared in terms of the NESS and
the MSE.

\begin{table*}
\centering
\begin{tabular}{|c|c|c||c|c|c|c|}
  \hline
  & mean NESS & std NESS & mean $MSE_1$ & mean $MSE_2$ & std $MSE_1$ & std $MSE_2$ \\
  \hline \hline
  PMC & 0.131 & 0.063 & 0.037 & $4.5 \times 10^{-3}$ & $135.4 \times 10^{-3}$& $591.1 \times 10^{-6}$ \\
  \hline
  NPMC temp & 0.937 & 0.059 & 0.019 & $3.3 \times 10^{-3}$ & $0.19 \times 10^{-3}$ & $5.71 \times 10^{-6}$ \\
  \hline
  NPMC clip & 0.937 & 0.060 & 0.019 & $3.3 \times 10^{-3}$ & $0.19 \times 10^{-3}$ & $5.68 \times 10^{-6}$ \\
  \hline\hline
  True posterior & - & - & 0.019 & $3.3 \times 10^{-3}$ & $0.18 \times 10^{-3}$ & $5.53 \times 10^{-6}$ \\
  \hline
\end{tabular}
\caption{Mean and standard deviation of the NESS, MSE of $\theta_1$
and MSE of $\theta_2$ for PMC, NPMC-temp and NPMC-clip.}
\label{GMM_table}
\end{table*}

The specific parameters of the plain PMC algorithm have been selected as
suggested in \cite{Cappe04} ($p=5$ scales, $\textbf{v}=[5, 2, 0.1,
0.05, 0.01]^\top$, $m=40$ samples per scale). A minimum of 1\% of $M$
of samples per scale has been kept as a baseline.

In the NPMC algorithm with tempering, the sequence of parameters $\gamma_\ell$,
has been obtained from the sigmoid function of the iteration index,
namely,
\begin{equation*}
\gamma_\ell = \frac{1}{1 + e^{-(\ell-5)}}, \quad \ell=1, \ldots, L.
\end{equation*}
With this choice of nonlinearity, the transformation of the weights is
practically eliminated after 10 iterations and the algorithm performs standard IS.

The NPMC algorithm with clipping has been simulated in its modified version, i.e.,
with the nonlinear transformation removed when the ESS reaches a
value of $M_{min}^{eff} = 100$. In this problem, the simulations
show that this occurs on average at the iteration $\ell=5$. The
parameter $M_T$ has been set to $M_T = M/4 = 50$ samples.

\begin{figure}[t]
\centering 
\includegraphics[width=0.49\textwidth]{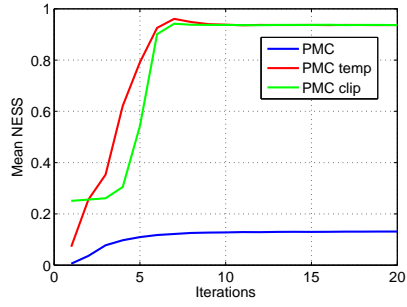}
\caption{ Evolution of the NESS for the standard PMC, NPMC with tempering and
NPMC with clipping.} \label{GMM_NESS}
\end{figure}

\begin{figure*}[t]
\centering 
\includegraphics[width=0.49\textwidth]{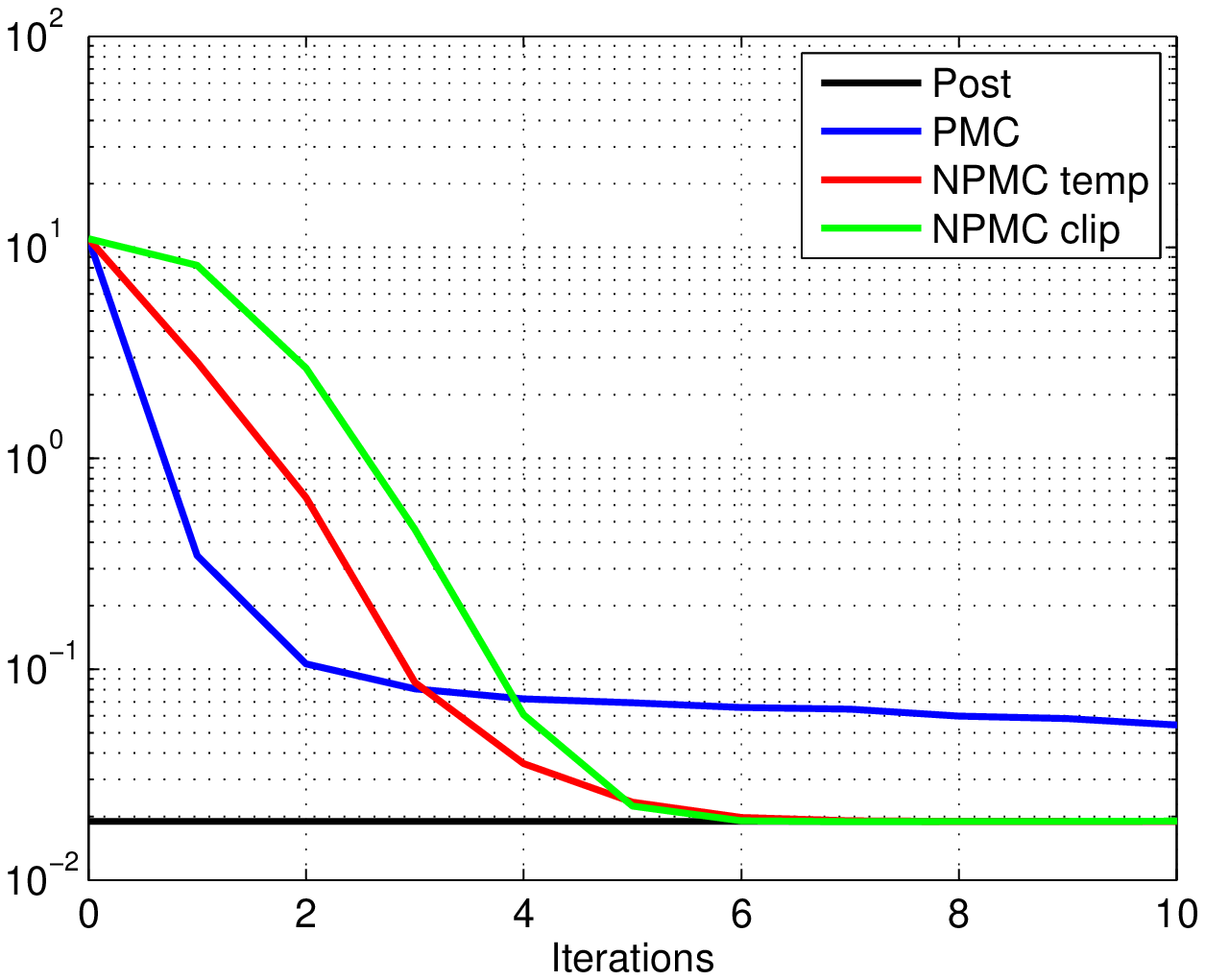}
\includegraphics[width=0.49\textwidth]{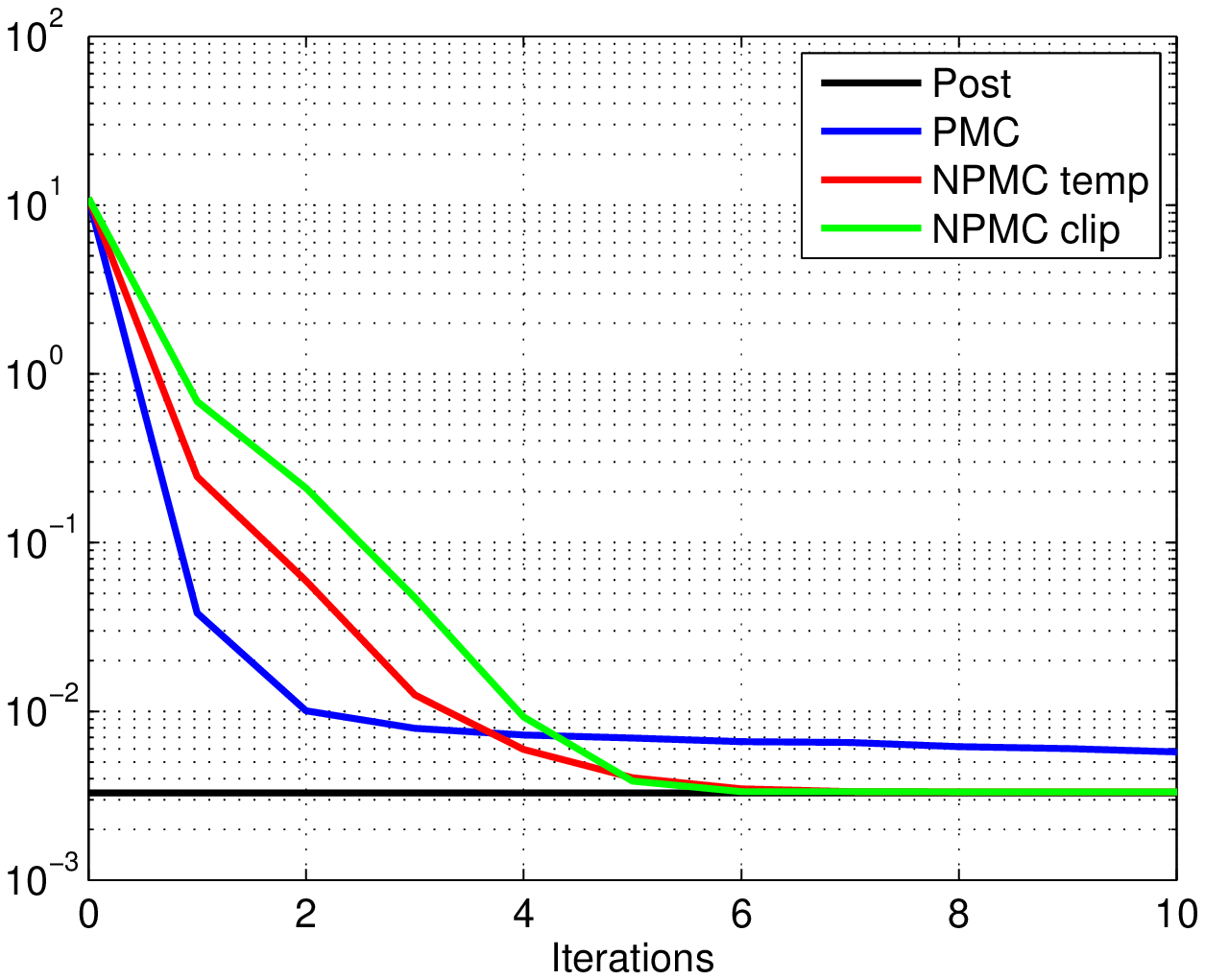}
\caption{ Evolution of the MSE of parameters $\theta_1$
(\emph{left}) and $\theta_2$ (\emph{right}) for the different PMC
schemes.} \label{GMM_MSE}
\end{figure*}

In Figures \ref{GMM_NESS} and \ref{GMM_MSE} numerical results
comparing the performance of the three algorithms (PMC, NPMC with tempering and
NPMC with clipping) are displayed. In Figure \ref{GMM_NESS} the evolution of
the average NESS along the iterations is depicted. It can be observed
that the original PMC scheme presents a low NESS, converging to
a value of\footnote{Note that the NESS ranges between 0 and 1.} 0.131. As a consequence of the degeneracy problem, the estimates
of the variables of interest obtained from this reduced set of
representative samples are often poor, presenting its
variance large fluctuations along the iterations. Since the proposal
density is updated based on the set of samples at the previous
iteration, the update procedure also suffers from the lack of useful
samples, leading to a very ``noisy'' convergence of the algorithm. On the contrary, the two NPMC schemes provide a smooth
convergence of the NESS to a value of 0.937 in about 10
iterations.


The degeneracy problem is most critical at the first iterations of
the PMC, when the proposal pdf does not usually fit the target
density. It is thus extremely hard to obtain a set of representative
samples (with non-negligible importance weights), which allow for a
consistent and robust proposal update. In Figure \ref{GMM_NESS}, the starting value of the
NESS corresponds to $M_T/M = 0.25$ for NPMC with clipping, 0.08 for
NPMC with tempering and around 0.01 for the standard PMC algorithm.

The mean and standard deviation of the NESS at the last iteration
are represented in Table \ref{GMM_table} for all three algorithms.
It can be checked that they present similar standard deviations,
while the standard PMC presents a considerably lower mean NESS than the
proposed methods.

In Figure \ref{GMM_MSE} the evolution along the iterations of the
MSE for $\theta_1$ (left) and $\theta_2$ (right) is represented for
the three algorithms. The minimum MSE of each parameter,
which has been numerically approximated from the posterior pdf
$p(\boldsymbol{\theta} | \textbf{y})$, is also shown for reference.
Again, it can be observed that the proposed NPMC schemes converge smoothly, 
reaching the minimum MSE in a low number of
iterations ($\ell=6$). On the other side, the original PMC does not
reach the minimum MSE with the given number of samples $M=200$.

However, the most outstanding difference in the performance of the
three algorithms is the extremely high variance of the MSE of the
original PMC scheme. Again due to degeneracy, this algorithm
presents a highly varying performance along the iterations and the
simulation runs. On the contrary, the proposed schemes reach the
minimum MSE, both in average and in standard deviation. The final
mean values, as well as the standard deviation of the MSE for
$\theta_1$ and $\theta_2$ are also shown in Table \ref{GMM_table}.

Moreover, the original PMC scheme hardly presents any adaptation
along most of the iterations and it allows for no parameterization
of the convergence speed. It reaches the final solution at the
second iteration, which is hardly improved for the rest of the
simulation. This behavior is not desirable in an iterative method,
and it suggests that the adaptation is not performed in a
appropriate way. On the contrary, the convergence speed of the
NPMC method with clipping may be adjusted by jointly selecting the parameters\footnote{Selecting a low value of $M_T$ leads to a
faster convergence to a worse estimate. On the contrary, an increase
in the value of $M_T$ slows down the convergence but the final
estimate has a smaller error. We have numerically
observed that a value for the ratio $M_T/M$ between 0.2 and 0.7
leads to good results.}
 $M$ and $M_T$,
and that of NPMC with tempering, by modifying the sequence $\gamma_\ell$.

\section{Example 2: A stochastic kinetic model}
\label{SKM_example}

In this section, the proposed nonlinear PMC method is applied to the
problem of estimating the hidden parameters of a simple stochastic
kinetic model (SKM), known as the predator-prey model. A SKM is a
multivariate continuous-time jump process modeling the interactions
among molecules, or species, that take place in chemical reaction
networks of biochemical and cellular systems
\cite{Wilkinson06,Boys08}.

Several MCMC schemes have been recently proposed to address the
problem of estimating the rate parameters in SKMs, \cite{Boys08},
\cite{Wilkinson11}, \cite{Milner06}, \cite{Golightly2011}. In
\cite{Milner06} the authors propose an approximation of the
likelihood based on the moment closure approximation of the
underlying stochastic process. In \cite{Golightly2011} a
likelihood-free particle MCMC scheme is applied to this problem. A
recently proposed work addressing this kind of problems is
\cite{Baragatti11}, where several MCMC schemes are proposed to
address likelihood-free scenarios.


We propose an alternative approach based on the NPMC method. In particular, we introduce a modification of the generic NPMC algorithm which
computes an approximation of the likelihood via a particle filter.
The proposed method is exact in the sense that it simulates from the
stochastic process in the model without further approximations.

\subsection{Predator-prey model}
\label{LV_model}

The Lotka-Volterra, or predator-prey, model is a simple SKM that
describes the time evolution of two species $x_1$ (prey) and $x_2$
(predator), by means of $K=3$ reaction equations:
\cite{Boys08,Volterra26}
\begin{equation*}
\begin{array}{cl}
  x_1 \stackrel{\theta_1} {\longrightarrow} 2 x_1 & \textrm{prey reproduction} \\
  x_1 + x_2 \stackrel{\theta_2} {\longrightarrow} 2 x_2 & \textrm{predator reproduction} \\
  x_2 \stackrel{\theta_3} {\longrightarrow} \emptyset & \textrm{predator death}
\end{array}
\end{equation*}

The $k$-th reaction takes place stochastically according to its
instantaneous rate $a_k(t) = \theta_k g_k \left( \textbf{x}(t)
\right)$, where $\theta_k > 0$ is the constant (yet random) rate parameter and $g_k
( \cdot )$ is a continuous function of the current state of the
system $\textbf{x}(t) = [x_1(t), x_2(t)]^\top$. We denote by
$x_1(t), x_2(t)$ the nonnegative, integer population of each species
at continuous time $t$. In this simple example, the instantaneous rates are of
the form
\begin{equation*}
a_1(t) = \theta_1 x_1(t), \; a_2(t) = \theta_2 x_1(t)x_2(t), \;
a_3(t) = \theta_3 x_2(t).
\end{equation*}

The waiting time to the next reaction is exponentially distributed
with parameter $a_0(t) = \sum_{k=1}^K a_k (t)$, and the probability
of each reaction type is given by $a_k(t)/a_0(t)$. We denote by $\textbf{x}$ the vector containing the population of
each species at the occurrence time of each reaction in a time
interval $t \in [0,T]$, i.e.,
\begin{equation*}
\textbf{x} = [ \textbf{x}^\top(t_1), \textbf{x}^\top(t_2), \ldots,
\textbf{x}^\top(t_R) ]^\top,
\end{equation*}
where $R$ is the total number of reactions and $t_i \in [0,T]$, $i=1, ..., R$, are the time instants at which the reactions occur.
Assuming that the entire vector $\textbf{x}$ is observed, the
likelihood function for the rate parameter vector
$\boldsymbol{\theta} = [\theta_1, \ldots, \theta_K]^\top$ can be
computed analytically \cite{Wilkinson06},
\begin{equation}
p (\textbf{x} | \boldsymbol{\theta}) = \prod_{k=1}^K p(\textbf{x} |
\theta_k) = \prod_{k=1}^K \theta_k^{r_k} \exp \left\{ - \theta_k
\int_0^T g_k \left( \textbf{x}(t) \right) dt \right\},
\label{eqLkhdWilkinson}
\end{equation}
where $r_k$ is the total number of reactions of type $k$ occurred in
the time interval $[0,T]$.

The structure of the likelihood function in (\ref{eqLkhdWilkinson}) allows the selection of a
conjugate prior distribution for the rate parameters in the form of
independent Gamma components, i.e.,
\begin{equation}
p(\boldsymbol{\theta}) = \prod_{k=1}^K p(\theta_k) =\prod_{k=1}^K
\mathcal{G} (\theta_k ; \, a_k, b_k),
\label{eqGammaPrior}
\end{equation}
where $a_k, b_k > 0$ are the scale and shape parameters of each
component, respectively. Thus, the posterior distribution $p (
\boldsymbol{\theta} | \textbf{x} ) = \prod_{k=1}^K p(\theta_k |
\textbf{x})$ may be also factorized into a set of independent Gamma
components
\begin{equation*}
\label{eq_post_theta_x} p(\theta_k | \textbf{x}) = \mathcal{G}
\left( \theta_k ; \; a_k + r_k, b_k + \int_0^T g_k \left(
\textbf{x}(t) \right) dt \right).
\end{equation*}
Therefore, in the complete-data scenario (in which $\textbf{x}$ is observed), exact inference
may be done for each rate constant $\theta_k$ separately. However,
making inference for complex, high-dimensional and discretely
observed SKMs (where $\textbf{x}$ is not fully available) is a
challenging problem \cite{Boys08}. 

Exact stochastic simulation of generic SKMs, and predator-prey
models in particular, can be carried out by the Gillespie algorithm
\cite{Gillespie77}. This procedure allows to draw samples from the
prior pdf of the populations, $p(\textbf{x} | \boldsymbol{\theta})$,
for arbitrary SKMs.

Although in this work we restrict ourselves to the predator-prey model, which is a simple but representative
example of SKM, the proposed algorithm may be equally applied to the
estimation of parameters of SKMs of higher complexity. 


\subsection{NPMC algorithm for SKMs}

We assume that a set of $N$ noisy observations of the populations of
both species are collected at regular time intervals of length
$\Delta$, that is, $\textbf{y}_n = \textbf{x}_n + \textbf{u}_n, \; n
= 1, \ldots, N$, where $\textbf{x}_n = [x_1(n\Delta),
x_2(n\Delta)]^\top$ and $\textbf{u}_n$ is a Gaussian noise component
with zero mean vector and covariance matrix $\sigma^2 \textbf{I}$.
We denote the complete vector of observations as $\textbf{y} = [
\textbf{y}_1^\top, \ldots, \textbf{y}_N^\top ]^\top$ with dimension
$2N \times 1$. Thus, the likelihood of the populations $\textbf{x}$
is given by
\begin{equation*}
p \left( \textbf{y} | \textbf{x} \right) = \prod_{n=1}^N \mathcal{N}
\left( \textbf{y}_n; \, \textbf{x}_n, \sigma^2 \textbf{I} \right).
\end{equation*}
The goal is to approximate the posterior distribution, with density $p
(\boldsymbol{\theta} | \textbf{y}) \propto p (\textbf{y} |
\boldsymbol{\theta} ) p (\boldsymbol{\theta})$, given the prior
pdf $p(\boldsymbol{\theta})$ and the likelihood
$p(\textbf{y} | \boldsymbol{\theta})$, using the NPMC method. 

In this particular problem, the observations $\textbf{y}$ are
related to the variables $\boldsymbol{\theta}$ through the random
vector $\textbf{x}$. Indeed, the likelihood of ${\boldsymbol
\theta}$ has the form
\begin{equation}
\label{int_y_theta} p(\textbf{y} | \boldsymbol{\theta}) = \int
p(\textbf{y} | \textbf{x}) p (\textbf{x} | \boldsymbol{\theta})
d\textbf{x} = E_{p(\textbf{x} | \boldsymbol{\theta})} \left[
p(\textbf{y} | \textbf{x}) \right],
\end{equation}
where $E_{p(\textbf{x} | \boldsymbol{\theta})} [\cdot]$ denotes
expectation with respect to the pdf in the subscript, and
$p(\textbf{y} | \textbf{x}, \boldsymbol{\theta}) = p(\textbf{y} |
\textbf{x})$, since the observations are independent of the
parameters $\boldsymbol{\theta}$ given the population vector
$\textbf{x}$.
As a consequence, the likelihood of the parameters cannot
be evaluated exactly. A set of likelihood-free techniques have
been recently proposed to tackle this kind of problems, which avoid
the need to evaluate the likelihood function \cite{Baragatti11}.
However, in this work we follow a different approach which consists
in computing an approximation of the likelihood of $p(\textbf{y} |
\boldsymbol{\theta})$.

In principle, it is possible to approximate the integral in
(\ref{int_y_theta}) as an average of the likelihoods $p (\textbf{y}
| \textbf{x}^{(i)})$ of a set $\{ \textbf{x}^{(i)}\}_{i=1}^I$ of
exact Monte Carlo samples from the density $p(\textbf{x} |
\boldsymbol{\theta})$, drawn using the Gillespie algorithm, that is,
\begin{equation*}
p(\textbf{y} | \boldsymbol{\theta}) \approx \frac{1}{I} \sum_{i=1}^I
p \left( \textbf{y} \left|\right. \textbf{x}^{(i)} \right).
\end{equation*}
This approach, however, is computationally intractable, because it
demands drawing a huge number of samples $I$ to obtain a useful
approximation of the likelihood $p(\textbf{y} |
\boldsymbol{\theta})$, since the probability of generating a
trajectory of populations $\textbf{x}^{(i)}$ similar to the
observations is extremely low.

To overcome this difficulty, we propose a simple approach based on
using a standard particle filter. An approximation
for the likelihood of the parameters $p( \textbf{y}
\left|\right. \boldsymbol{\theta}^{(i)} )$ may be recursively
obtained based on the particle filter approximation of the posterior
of the populations $p ( \textbf{x} \left|\right. \textbf{y},
\boldsymbol{\theta}^{(i)} )$. Indeed, the likelihood of a sample
$\boldsymbol{\theta}^{(i)}$ can be recursively factorized as
\begin{equation*}
p \left( \textbf{y} \left|\right. \boldsymbol{\theta}^{(i)} \right)
= \prod_{n=1}^N p \left( \textbf{y}_n \left|\right.
\textbf{y}_{1:n-1}, \boldsymbol{\theta}^{(i)} \right)
\end{equation*}
where each of the terms in the product (with fixed $\boldsymbol{\theta} = {\boldsymbol{\theta}}^{(i)}$) can be approximated via 
particle filtering (see, e.g., \cite{Djuric10} or \cite{Maiz12}).

The NPMC method for this application is exact in the sense that it uses exact samples
$\textbf{x}^{(i)}$ from the stochastic model generated via the
Gillespie algorithm, and does not perform any approximation of the
stochastic process. Alternative approaches have been proposed which
employ a diffusion approximation. However, this approximation has
not shown to introduce any improvement to our algorithm.

To summarize, the standard importance weights for this
application have the form
\begin{equation*}
w_\ell^{(i)*} \propto \frac{ \hat p \left( \textbf{y} \left|\right.
\boldsymbol{\theta}_\ell^{(i)} \right) p \left(
\boldsymbol{\theta}_\ell^{(i)} \right) }{q_\ell \left(
\boldsymbol{\theta}_\ell^{(i)} \right)},
\end{equation*}
where the likelihood $\hat p( \textbf{y} | \boldsymbol{\theta}_\ell^{(i)})$  is approximated using a particle filter. The
prior is a product of $K$ independent Gamma components as in Eq. (\ref{eqGammaPrior}) and the
proposal is a multivariate normal pdf, as shown in Table \ref{tNPMCalgorithm}.

Let us finally remark that, while in this work we have focused on the complete-observation scenario, the proposed
method is also valid in the partially observed case, when only a subset of the species can be observed.

\subsection{Computer simulation results}
\label{Num_results}

We have applied the NPMC algorithm to the problem
of estimating the posterior pdf of the constant rate parameters
vector $\boldsymbol{\theta}$ in a simple predator-prey model. The
true vector of parameter rates which we aim to estimate has been set
to
\begin{equation*}
\boldsymbol{\theta} = [0.5, 0.0025, 0.3]^\top.
\end{equation*}

A realization of the populations $\textbf{x}$ has been generated
from the prior distribution $p(\textbf{x} | \boldsymbol{\theta})$
with initial populations  $\textbf{x}(0) = [ 71, 79 ]^\top$,
and a total length of $T=40$. The observation vector $\textbf{y}$
has been generated with an observation period $\Delta = 1$ and a
Gaussian noise variance $\sigma^2 = 100$.

\begin{figure}[h]
\centering 
\includegraphics[width=0.5\textwidth]{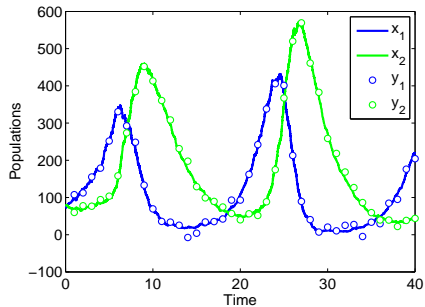}
\caption{Real populations ($\textbf{x}$) and discrete-time noisy
observations ($\textbf{y}$) in a predator-prey model.}
\label{Populations}
\end{figure}

Figure \ref{Populations} depicts the time evolution of the true
populations of both species $\textbf{x}$, and the corresponding
discrete-time noisy observations $\textbf{y}$. The autoregulatory
behavior of the model can be clearly observed on the graph.

The parameters of the prior Gamma distribution
$p(\boldsymbol{\theta})$ have been chosen such that the corresponding
mean matches the real value of $\boldsymbol{\theta}$ and the
marginal standard deviations are $[1.25, 0.0065, 0.77]^\top$. These
priors cover an order of magnitude either side of the true value,
which represents a vague prior knowledge.


The target density for the NPMC algorithm is the posterior pdf, $\pi(\boldsymbol{\theta}) = p( \boldsymbol{\theta} | \textbf{y} )$. Despite the low dimension of this problem ($K=3$), the importance
weights of the PMC scheme present severe degeneracy, partially due
to the likelihood approximation. In the simulations, we have observed that the importance weights present such
degeneracy, that, for the NPMC scheme with tempering, the coefficient
$\gamma_\ell$ must take extremely low values (about $10^{-4}$) at
the first iterations in order to balance the weights and obtain a
reasonable number of effective samples. Besides, the sequence
$\gamma_\ell$ is hard to design a priori, and it does not guarantee
that the resulting sample set leads to accurate estimates. For
that reasons, in this example we have restricted our attention to the hard clipping technique, which does not
require the fitting of any parameters and guarantees a baseline
ESS. Recall that the NPMC scheme with hard clipping only requires the selection of a unique
parameter $M_T$, which allows to adjust the convergence performance
of the algorithm, and does not require a precise fitting. In this example, we have chosen $M_T = M/5$, with a total number of samples $M=500$ and $L=10$ iterations.

In order to be able to compare the MSE of parameters $\theta_k$, $k=1, 2, 3$, that
take very different values, we additionally define normalized MSEs (NMSE) as
\begin{equation*}
NMSE_k = \frac{1}{M} \sum_{i=1}^M \left(\frac{
\tilde{\theta}_k^{(i)} - \theta_k }{\theta_k}\right)^2 =
\frac{MSE_k}{\theta_k^2}, \quad k=1, 2, 3,
\end{equation*}
and a global error figure as the average NMSE, namely
\begin{equation*}
NMSE = \frac{1}{K} \sum_{k=1}^K NMSE_k.
\end{equation*}

We have carried out two different experiments. In both experiments
we have performed $P=100$ independent simulation runs, with the same
true parameters $\boldsymbol{\theta}$ and initial populations
$\textbf{x}(0)$, but different (independent) population and observations vectors
$\textbf{x}$ and $\textbf{y}$ in each run.

In the first experiment we considered the same number of
samples $M=500$ in all the simulation runs of the NPMC algorithm and
computed the NESS and the NMSE along the iterations. We observed that a 70\% of the simulation runs converged to yield accurate estimates (with low NMSE), while in a 30\% of the cases, the algorithm produced biased estimates, with considerably higher NMSE and low final NESS (below a 0.3).

\begin{figure*}[t]
\centering \hspace{-1cm}
\includegraphics[width=0.55\textwidth]{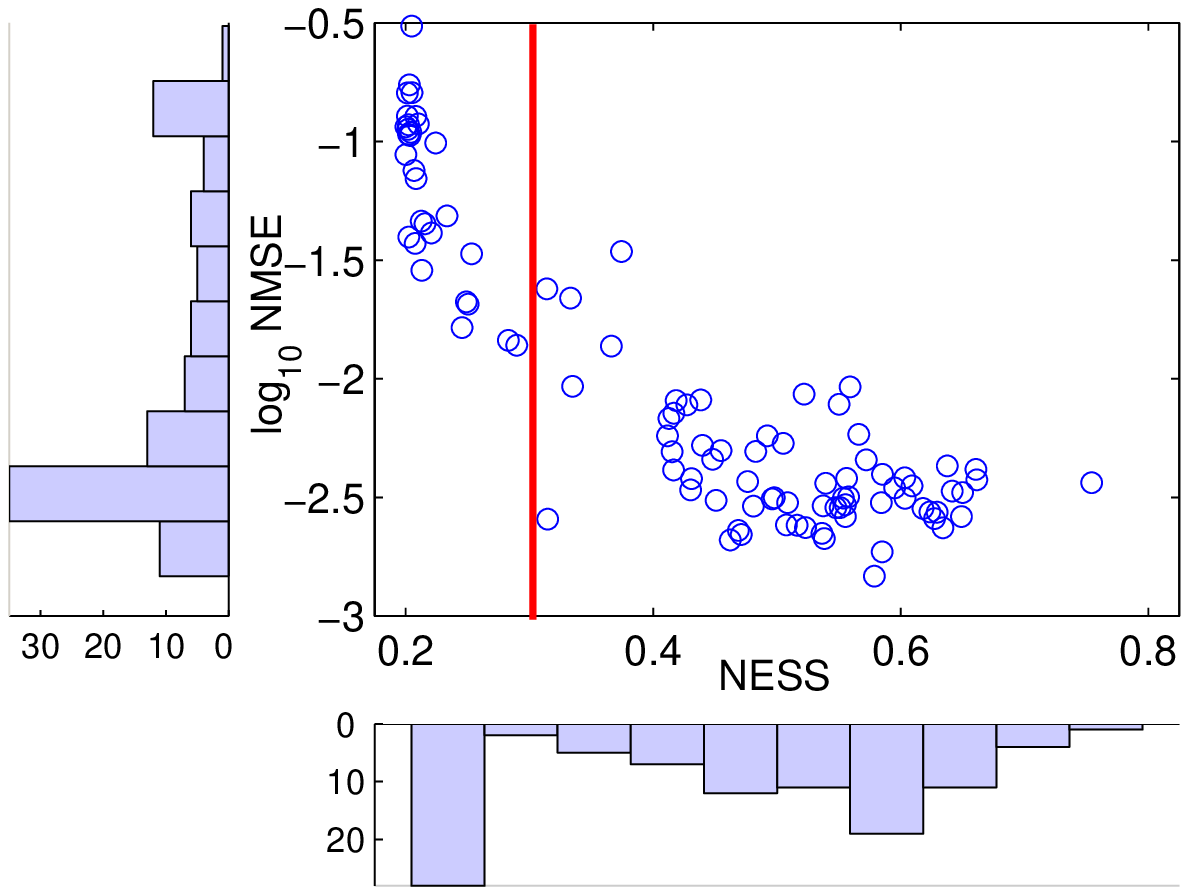}
\hspace{-1.1cm}
\includegraphics[width=0.55\textwidth]{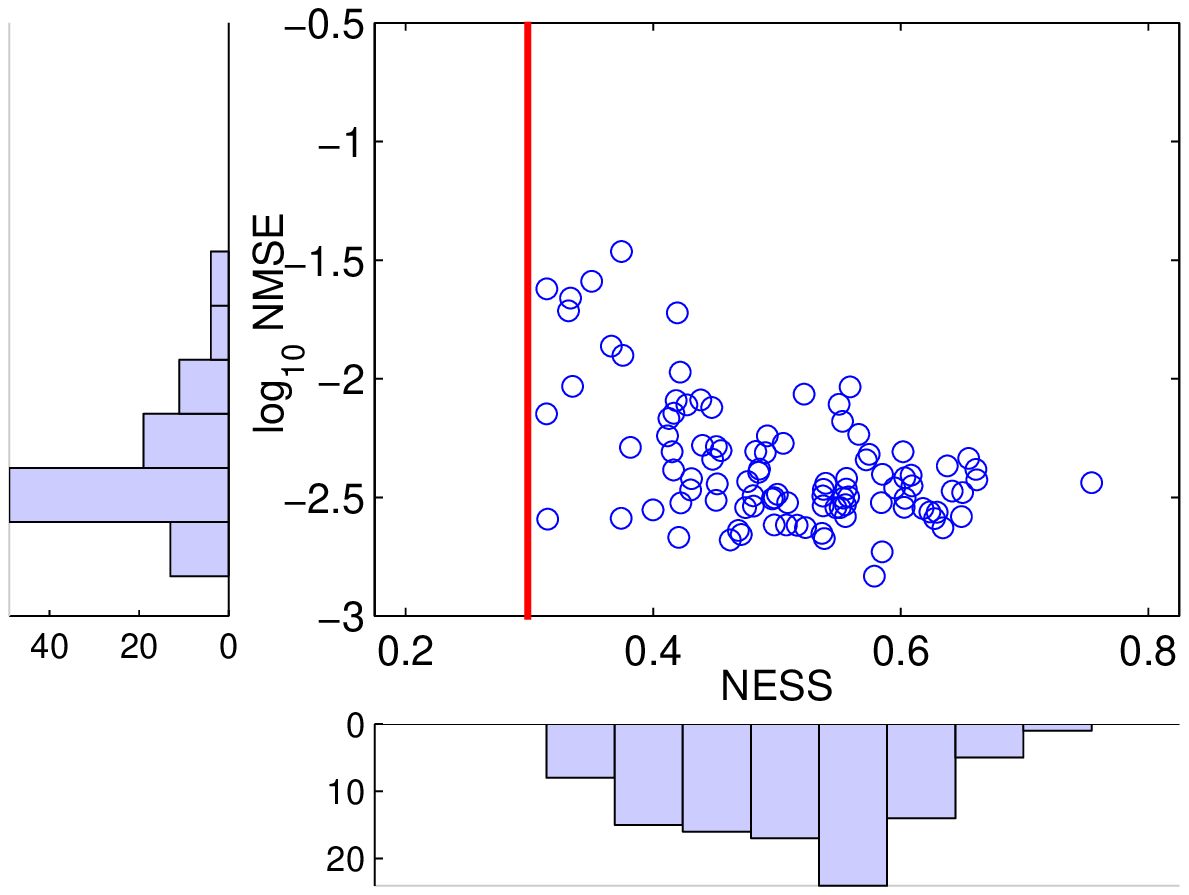}
\caption{{\em Left:} Average value of $\log_{10}(\mbox{NMSE})$ versus the NESS after the last iteration ($L=10$) for each
simulation run, with $M=500$ samples. The histogram of each variable is also represented. 
{\em Right:} Average value of $\log_{10}(\mbox{NMSE})$ versus the NESS after the last iteration ($L=10$) for each
simulation run, with $M=500$ for 70\% of the simulation runs,
$M=1000$ for the 28\% of the simulation runs and $M=2000$ for 8\% of the simulation runs.} \label{SKM_scatter}
\end{figure*}

\begin{figure*}[t]
\centering 
\includegraphics[width=0.49\textwidth]{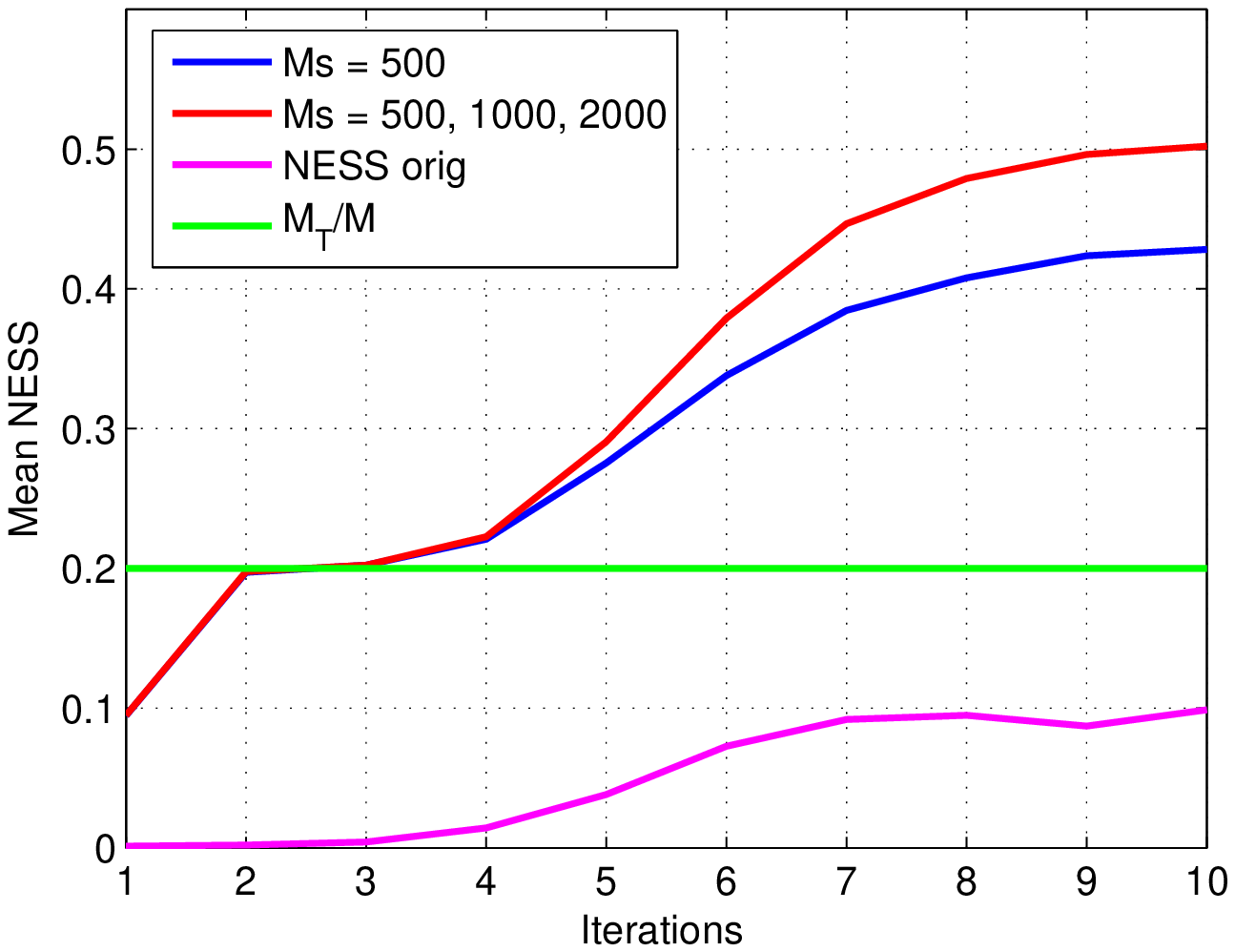}
\includegraphics[width=0.49\textwidth]{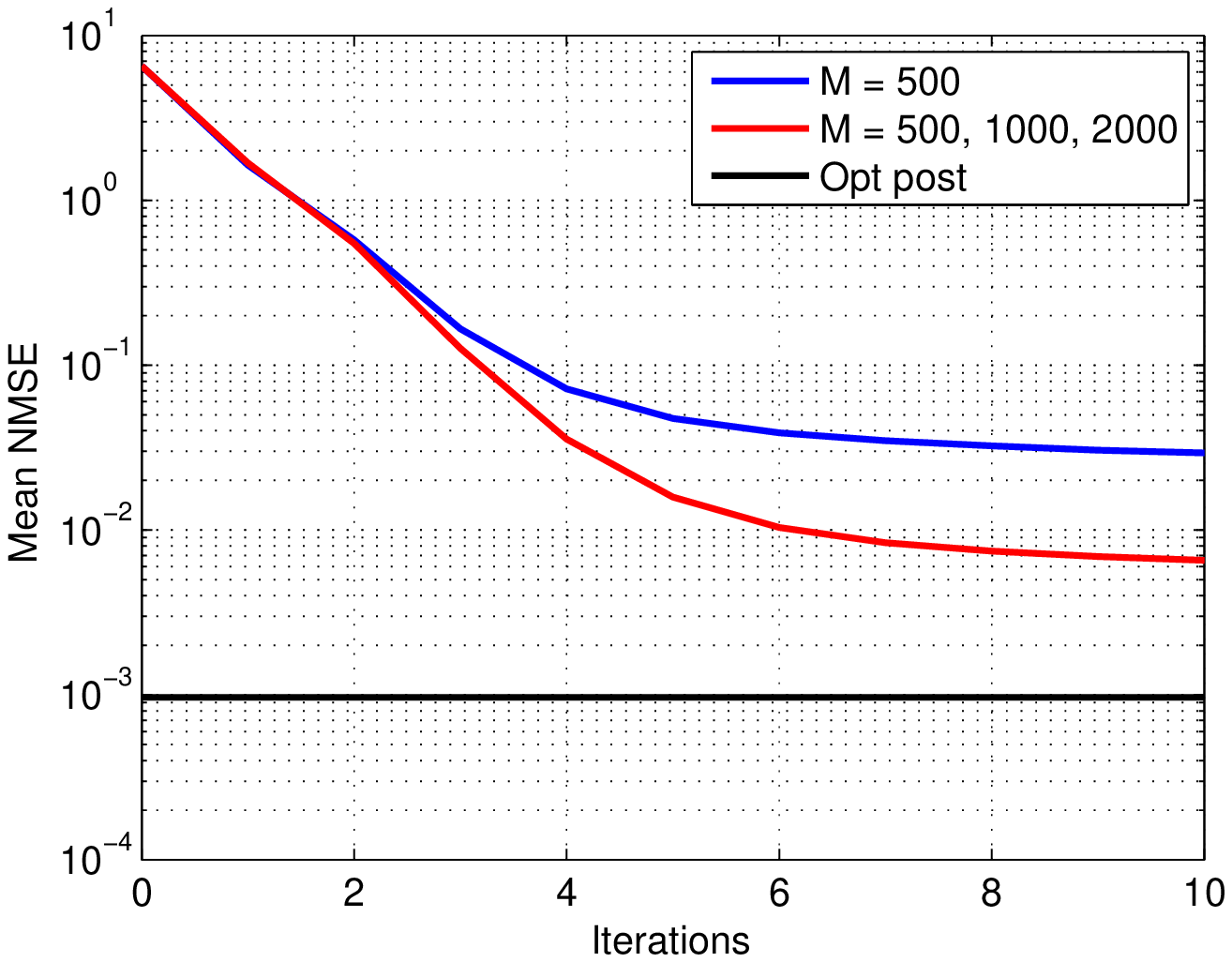}
\caption{{\em Left:} Evolution of the mean NESS, before and after
clipping. The NESS before clipping (computed for the standard weights) is labeled {\em NESS orig}. {\em Right:} Evolution of the mean MSE. The lower bound is
obtained from the optimal posterior $p(\boldsymbol{\theta} |
\textbf{x})$ (labeled {\em Opt post} in the figure).} \label{SKM_NESS}
\end{figure*}

In Figure \ref{SKM_scatter} (left) the final values of the logarithm of the NMSE
versus the NESS obtained for each simulation run are depicted, together with
the histograms of both the $\log_{10}(\mbox{NMSE})$ and the NESS. This plot reveals how low
NESS values correspond to high NMSE, and viceversa. This empirical
relationship is relevant because allows to detect, in practice, whether the algorithm has converged
properly or not.

The second experiment consisted in repeating the simulation of
the algorithm in those cases which did not yield accurate estimates with $M=500$
samples, but increasing the number of samples to $M=1000$. For this new experiment, we 
considered that the algorithm had converged when the final NESS was greater than 0.3. Out of the repeated simulations, 8 cases still
did not converge and the simulations were repeated with $M=2000$. Finally, all the
simulation runs obtained a final NESS$>0.3$ with a number of
particles $M=500$ for the 70\% of the cases, $M=1000$ for the 22\%
of the cases and $M=2000$ for the 8\% of the cases.

In Figure \ref{SKM_scatter} (right) the final values of $\log_{10}(\mbox{NMSE})$
versus the NESS are plotted after repeating the simulations with
$M=1000$ (22\% of the runs) and $M=2000$ (8\% of the runs). It
can be observed that the points with high NMSE are removed and that
most points concentrate in the region of high NESS and low NMSE.
Also the histograms of NMSE and NESS show how their mean values
significantly improve.

\begin{table*} \centering
\begin{tabular}{|c||c|c|c||c|c|c||c|}
  \hline
  & MSE $\theta_1$ & MSE $\theta_2$ & MSE $\theta_3$ & NMSE $\theta_1$ & NMSE $\theta_2$ & NMSE $\theta_3$ & mean NMSE \\
  \hline \hline
  Prior & $1.56$ & $4.42 \times 10^{-5}$ & $0.5929$ & $6.25$ & $6.76$ & 6.588 & 6.533 \\
  \hline\hline
  NPMC clip (1) & $9.7 \times 10^{-3}$ & $53.0 \times 10^{-9}$ & $36.8 \times 10^{-4}$ & 0.03878 & 0.008484 & 0.04089 & 0.02938 \\
  \hline
  NPMC clip (2) & $1.4 \times 10^{-3}$ & $23.46 \times 10^{-9}$ & $6.38 \times 10^{-4}$ & 0.005618 & 0.003754 & 0.007092 & \textbf{0.005488} \\
  \hline\hline
  Optimal posterior & $0.26 \times 10^{-3}$ & $5.53 \times 10^{-9}$ & $0.881 \times 10^{-4}$ & $0.00103$ & $0.0008862$ & 0.0009796 & 0.0009654 \\
  \hline
\end{tabular}
\caption{Final mean MSE and NMSE for the parameters $\theta_1$,
$\theta_2$ and $\theta_3$ in experiments 1 and 2. The prior and
optimal posterior values are included for comparison.}
\label{tabla2}
\end{table*}

\begin{figure*}[t]
\centering \hspace{-0.2cm}
\includegraphics[width=0.35\textwidth]{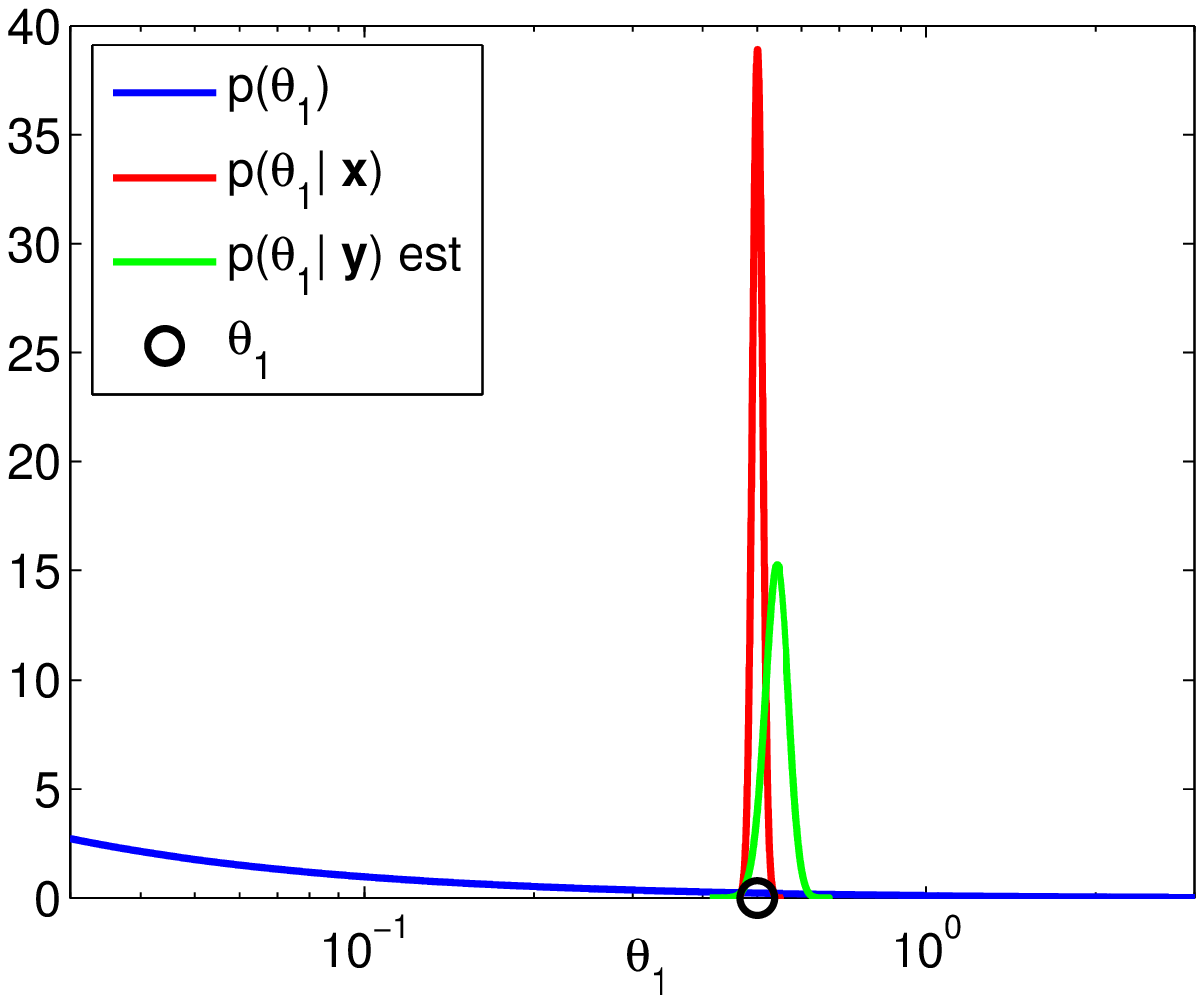}
\hspace{-0.6cm}
\includegraphics[width=0.35\textwidth]{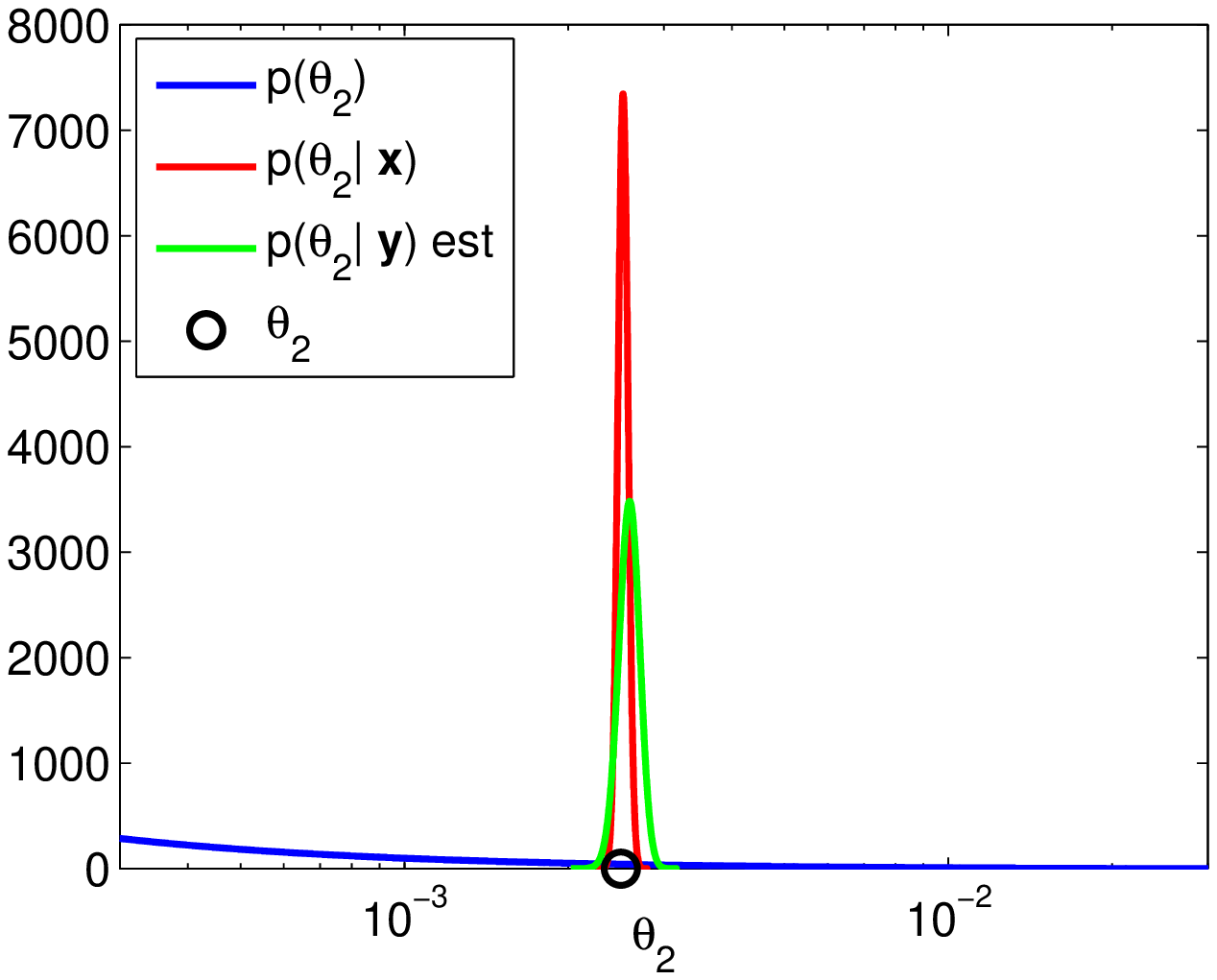}
\hspace{-0.6cm}
\includegraphics[width=0.35\textwidth]{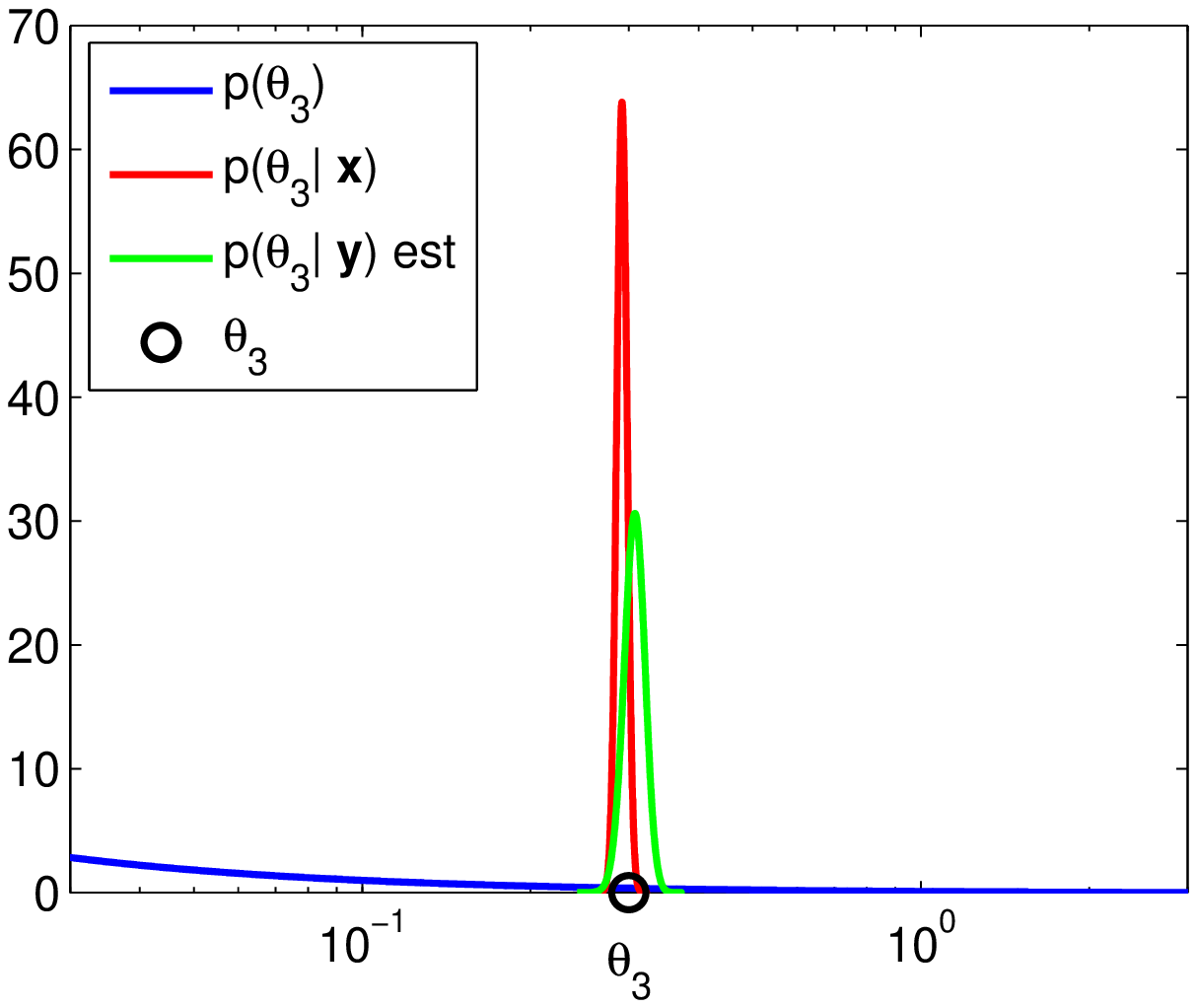}
\caption{Example of pdfs of interest of a particular simulation run
with final NMSE of 0.0053 (close to the average NMSE): marginal
priors $p(\theta_k)$, optimal posteriors $p(\theta_k|\textbf{x})$,
estimated posteriors $\hat{p}(\theta_k | \textbf{y})$ and true
values $\theta_k$ for $k=1$ (\emph{left}), $k=2$ (\emph{central})
and $k=3$ (\emph{right}).} \label{Densities}
\end{figure*}

Figure \ref{SKM_NESS} shows the evolution of the mean NESS and NMSE
along the iterations in both experiments. In solid blue lines, the
results regarding the first experiment are plotted ($M=500$ for all
simulation runs). In solid red lines, the results of the second
experiment are represented ($M=500$, $M=1000$ and $M=2000$).

In Figure \ref{SKM_NESS} (left) the evolution of the NESS,
$M_\ell^{neff}$, after the clipping procedure in both experiments is
represented. In both cases it starts from an initial value of
$M_T/M$ and reaches a final value of about 0.43 in the first case
and 0.5 in the second case\footnote{Note that the NESS increases beyond the effect of the clipping procedure,
which indicates that the iterative scheme is able to generate more
representative samples as the algorithm converges.}. Also the NESS before clipping (i.e., for the standard weights) is
represented in this graph, which takes significantly lower values,
but presents a significant increase after the first iterations. In this
example, the clipping procedure has been performed along all the
iterations because of the severe degeneracy of the weights. This is
probably due to the approximation in the computation of the
likelihood $p(\textbf{y}|\boldsymbol{\theta})$, which leads to
additional fluctuations in the values of the importance weights,
increasing the effects of degeneracy.


Figure \ref{SKM_NESS} (right) shows the evolution of the NMSE
along the iterations, as well as the corresponding lower bound given
by the optimal marginal posterior with complete data, $p(\theta_k | \textbf{x})$, in
both experiments. The starting value at $\ell=0$ corresponds to the
vague knowledge described by the prior distribution. It can be seen
that the NMSE of the NPMC algorithms with clipping smoothly decreases until a
final value close to the lower bound, in a low number of iterations.

Table \ref{tabla2} summarizes the obtained results in terms of MSE
and NMSE for each parameter $\theta_k$ for both experiments. It can
be observed that the prior distribution provides vague knowledge of
the variables of interest, with a high MSE and NMSE. The NMSE allows
to compare the error values of the three parameters, even when
$\theta_2$ has a very low value with respect to $\theta_1$ and
$\theta_3$.

In the first experiment we obtain a remarkable reduction in MSE and
NMSE with respect to the prior. However, increasing the number of
samples in a fraction of cases allows to further reduce the
estimation error. The MSE and NMSE of the optimal marginal
posteriors are also shown for comparison. Note, however, that this is an
unreachable performance bound, since it is obtained when the whole vector
$\textbf{x}$ is observed.





In order to illustrate the quality of the obtained approximations,
Figure \ref{Densities} depicts the final estimate of the marginal
posteriors $p(\theta_k | \textbf{y})$, together with the
corresponding prior $p(\theta_k)$ and the optimal posterior in the
complete-data scenario $p(\theta_k | \textbf{x})$. It can be
observed that the algorithm provides a good approximation of the


\section{Conclusion}
\label{Conclusion}

We have addressed the problem of approximating posterior probability
distributions by means of random samples. A recently proposed approach
to tackle this problem is the population Monte Carlo method, that
consists in iteratively approximating a target distribution via an
importance sampling scheme. The main limitation of this algorithm is
that is presents severe degeneracy of the importance weights as the
dimension of the model, $K$, and/or the number of observations, $N$, increase. This leads to a highly varying
number of effective samples and inaccurate estimates, unless the
number of samples is extremely high (which makes the method computationally prohibitive).

We propose to apply a simple procedure in order to guarantee a prescribed ESS and a smooth and robust convergence. It consists in applying nonlinear transformations to the standard importance weights in order to reduce their fluctuations and thus avoid degeneracy. In order to illustrate the application of the proposed technique, we have applied it to two examples of different complexity. The first example is a simple GMM, which allows to get insight of the performance of the standard PMC scheme, the degeneracy problem
and the behavior of the proposed algorithm.

We have tackled the problem of estimating the set of constant (and random) rate parameters of a stochastic kinetic model. Even for the relatively simple predator-prey model that we have studied, this is significantly more complex than the GMM example. We have shown extensive simulation results that show how the proposed NPMC scheme can greatly improve the performance of the standard method.  

The convergence of standard PMC algorithms is often justified by the asymptotic convergence of IS (with respect to the number of samples). The NPMC scheme modifies the importance weights and, hence, the standard theory of IS cannot be applied directly. To address this difficulty, we have analyzed the convergence of the approximations of integrals computed using transformed importance weights and proved that they converge a.s., similar to the results available for standard IS.

\begin{acknowledgements}

E. K.  acknowledges the support of \textit{Ministerio de
Educaci\'on} of Spain (\textit{Programa de Formaci\'on de
Profesorado Universitario}, ref. AP2008-00469). This work has been
partially supported by {\em Ministerio de Econom\'{\i}a y Competitividad} of
Spain (program Consolider-Ingenio 2010 CSD2008-00010 COMONSENS and
project DEIPRO TEC2009-14504-C02-01).

\end{acknowledgements}

\appendix

\section{Proof of Lemma \ref{L1}}
\label{App_1}

As a first step, we seek a tractable upper bound for the difference
\begin{equation}
\left| 
	(f,\bar \pi^M) - (f,\check \pi^M)
\right| = \left|
	\sum_{i=1}^M f({\boldsymbol{\theta}}^{(i)}) \left(
		\bar w^{(i)} - \check w^{(i)}
	\right)
\right|,
\label{eqDif-1}
\end{equation}
where
\begin{eqnarray}
\bar w^{(i)} &=& \frac{
	(\varphi^M \circ g)({\boldsymbol{\theta}}^{(i)})
}{
	\sum_{j=1}^M (\varphi^M \circ g)({\boldsymbol{\theta}}^{(j)})
} \quad \mbox{and} \label{eqWeights-1} \\
\check w^{(i)} &=& \frac{
	(\varphi^M \circ g)({\boldsymbol{\theta}}^{(i)})
}{
	\sum_{j=1}^M g({\boldsymbol{\theta}}^{(j)})
}, \label{eqWeights-2}
\end{eqnarray}
and $(\varphi^M \circ g)(\boldsymbol{\theta}) = \varphi^M( g(\boldsymbol{\theta}) )$ denotes the composition of the functions $\varphi^M$ and g. Moreover, the constants in the denominators of the weights can be written as integrals with respect to the random measure 
\begin{equation}
q^M(d\boldsymbol{\theta}) = \frac{1}{M} \sum_{j=1}^M \delta_{{\boldsymbol{\theta}}^{(j)}}(d\boldsymbol{\theta}),
\label{eqPriorRandomMeasure}
\end{equation}
namely, 
\begin{equation}
\sum_{j=1}^M (\varphi^M \circ g)({\boldsymbol{\theta}}^{(j)}) = M(\varphi^M \circ g, q^M) 
\label{eqInt-1}
\end{equation}
and
\begin{equation}
\sum_{j=1}^M g({\boldsymbol{\theta}}^{(j)}) = M(g,q^M). 
\label{eqInt-2}
\end{equation}
Substituting (\ref{eqWeights-1}), (\ref{eqWeights-2}), (\ref{eqInt-1}) and (\ref{eqInt-2}),  into (\ref{eqDif-1}) yields, after straightforward manipulations,
\begin{equation}
\left| 
	(f,\bar \pi^M) - (f,\check \pi^M)
\right| = \left|
	\frac{1}{M}Ê\sum_{i=1}^M f({\boldsymbol{\theta}}^{(i)}) (\varphi^M \circ g)({\boldsymbol{\theta}}^{(i)}) 
	\frac{
		(g,q^M) - (\varphi^M \circ g, q^M)
	}{
		(\varphi^M \circ g, q^M) (g,q^M)
	}
\right|. \label{eqEquality-0}
\end{equation}
A useful upper bound for the difference of integrals follows quite easily from (\ref{eqEquality-0}). In particular, note that $| f({\boldsymbol{\theta}}^{(i)}) (\varphi^M \circ g)| \le \| f \|_\infty \| g \|_\infty$, since $f,g \in B(\mathbb{R}^K)$ and $\varphi^M \circ g \le g$, while the latter inequality also implies that $(\varphi^M \circ g, q^M) (g,q^M) \ge (\varphi^M \circ g, q^M)^2$. Also note that, from the definition of $\varphi^M$,  
\begin{eqnarray}
| (g,q^M) - (\varphi^M \circ g, q^M) | &\le& \frac{1}{M} \sum_{k=1}^{M_T} | g({\boldsymbol{\theta}}^{(i_k)}) - (\varphi^M \circ g) ({\boldsymbol{\theta}}^{(i_k)}) | \nonumber\\
&\le& \frac{
	2 M_T \| g \|_\infty
}{
	M
}.
\label{eqAlmendruco-1}
\end{eqnarray}
As a result, we obtain 
\begin{equation}
\left| 
	(f,\bar \pi^M) - (f,\check \pi^M)
\right| \le \frac{
	2 \| f \|_\infty \| g \|_\infty^2 M_T 
}{
	M (\varphi^M \circ g, q^M)^2
}. \label{eqUpperBound}
\end{equation}

Let $c_1 > 0$ be some arbitrary real constant. From (\ref{eqUpperBound}),
\begin{equation}
\mathbb{P}\left\{
	\left| 
	(f,\bar \pi^M) - (f,\check \pi^M)
\right| \le \frac{
	M_T 
}{
	M
} c_1
\right\} \ge \mathbb{P}\left\{
	\frac{
		2 \| f \|_\infty \| g \|_\infty^2 M_T 
	}{
		M (\varphi^M \circ g, q^M)^2
	} \le 
	\frac{
		M_T 
	}{
		M
	} c_1
\right\}. 
\label{eqCharranga}
\end{equation}
If we choose 
\begin{equation}
c_1 = \frac{
	2 \| f \|_\infty \| g \|_\infty^2 
}{
	\left(
		\frac{1}{a} - \frac{1}{\sqrt{2}}
	\right)^2 (g,q)^2
},
\label{eqDef_c_1}
\end{equation}
where $1 < a < \sqrt{2}$, then substituting (\ref{eqDef_c_1}) into the right hand side of (\ref{eqCharranga}) yields
\begin{eqnarray}
\mathbb{P}\left\{
	\frac{
		2 \| f \|_\infty \| g \|_\infty^2 M_T 
	}{
		M (\varphi^M \circ g, q^M)^2
	} \le 
	\frac{
		M_T 
	}{
		M
	} c_1
\right\} &=& \mathbb{P}\left\{
	(\varphi^M \circ g, q^M)^2 \ge \left(
		\frac{1}{a} - \frac{1}{\sqrt{2}}
	\right)^2 (g,q)^2
\right\} \nonumber \\
&=& \mathbb{P}\left\{
	(\varphi^M \circ g, q^M)  - \frac{1}{a} (g,q)
	\ge -\frac{1}{\sqrt{2}} (g,q)
\right\}, \nonumber \\
&=& \mathbb{P}\left\{
	M(\varphi^M \circ g, q^M)  - \frac{M}{a} (g,q)
	\ge -\frac{M}{\sqrt{2}} (g,q)
\right\},
\label{eqCharranga-2}
\end{eqnarray}
where the second equality holds because $\frac{1}{a}-\frac{1}{\sqrt{2}}>0$.

Next, consider the expectations $E_q[ (\varphi^M \circ g, q^M) ]$ and $E_q[ (g, q^M) ] = (g,q)$. Since, $| (\varphi^M \circ g, q^M) - (g, q^M) | \le 2M_T\|g\|_\infty/M$ (see Eq. (\ref{eqAlmendruco-1})), it follows that
\begin{eqnarray}
\left|
	E_q\left[ (\varphi^M \circ g, q^M) \right] - E_q\left[ (g, q^M) \right]
\right| &\le& E_q\left[
	| (\varphi^M \circ g, q^M) - (g, q^M) |
\right] \nonumber\\
&\le& \frac{2M_T\|g\|_\infty}{M}. \nonumber
\end{eqnarray} 
Therefore, since we have assumed that $\lim_{M\rightarrow \infty} \frac{M_T}{M} = 0$ and $g>0$, there exists $M_a$ such that, for all $M > M_a$,
\begin{equation}
E_q\left[ (\varphi^M \circ g, q^M) \right] > \frac{1}{a} E_q\left[ (g,q^M) \right] = \frac{1}{a}(g,q),
\label{eqAlmendruco-2}
\end{equation} 
and combining Eq. (\ref{eqCharranga-2}) with the inequality (\ref{eqAlmendruco-2}) we obtain that
\begin{eqnarray}
\mathbb{P}\left\{
	\frac{
		2 \| f \|_\infty \| g \|_\infty^2 M_T 
	}{
		M (\varphi^M \circ g, q^M)^2
	} \le 
	\frac{
		M_T 
	}{
		M
	} c_1
\right\} \ge && \nonumber \\
\mathbb{P}\left\{
	M(\varphi^M \circ g, q^M)  - M E_q\left[ (\varphi^M \circ g, q^M) \right]
	\ge -\frac{M}{\sqrt{2}} (g,q)
\right\}. && \label{eqAhuevo}
\end{eqnarray}

Since $M(\varphi^M \circ g, q^M) = \sum_{i=1}^M \varphi^M( g( {\boldsymbol{\theta}}^{(i)} ) )$ is the sum of $M$ independent and bounded random variables, each of them taking values within the interval $(0,\| g \|_\infty)$, it is straightforward to apply Hoeffding's tail inequality \cite{Hoeffding63} to obtain a lower bound on (\ref{eqAhuevo}), namely
\begin{equation}
\mathbb{P}\left\{
	M(\varphi^M \circ g, q^M)  - M E_q\left[ (\varphi^M \circ g, q^M) \right]
	\ge -\frac{M}{\sqrt{2}} (g,q)
\right\} \ge 1 - \exp\left\{
	-\frac{(g,q)^2}{\| g \|_\infty^2}M.
\right\}.  
\label{eqAhuevo-2}
\end{equation}

Substituting (\ref{eqAhuevo-2}) back into (\ref{eqAhuevo}), (\ref{eqCharranga-2}) and (\ref{eqCharranga}) yields the 
desired result,
\begin{equation}
\mathbb{P}\left\{
	\left| 
	(f,\bar \pi^M) - (f,\check \pi^M)
\right| \le \frac{
	M_T 
}{
	M
} c_1
\right\} \ge 1 - \exp\left\{
	-c_1' M
\right\},
\label{eqL1Part1}
\end{equation}
with $c_1' = \frac{(g,q)^2}{\| g \|_\infty^2}$. 


\section{Proof of Lemma \ref{L2}}
\label{App_2}

The argument is similar to that of the proof of Lemma \ref{L1}. Recalling Eqs. (\ref{eqWeights-2}), (\ref{eqPriorRandomMeasure}) and (\ref{eqInt-2}) in Appendix \ref{App_1} as well as the form of the standard normalized weights, 
\begin{equation*}
w^{(i)} = \frac{
	g({\boldsymbol{\theta}}^{(i)})
}{
	M(g,q^M)
},
\end{equation*}
it is straightforward to show that
\begin{equation*}
\left|
	(f,\check \pi^M) - (f,\pi^M)
\right| = \frac{
	1
}{
	M(g,q^M)
} \left|
	\sum_{k=1}^{M_T} f({\boldsymbol{\theta}}^{(i_k)})\left(
		(\varphi^M \circ g)({\boldsymbol{\theta}}^{(i_k)}) - g({\boldsymbol{\theta}}^{(i_k)})
	\right)
\right|,
\end{equation*}
which readily yields the upper bound
\begin{equation}
\left|
	(f,\check \pi^M) - (f,\pi^M)
\right| \le \frac{
	2 \|f\|_\infty \|g\|_\infty M_T
}{
	M (g,q^M)
}.
\label{eqUpperBound-L2}
\end{equation}

Let $c_2 > 0$ be some arbitrary real constant. From (\ref{eqUpperBound-L2}),
\begin{equation}
\mathbb{P}\left\{
	\left|
		(f,\check \pi^M) - (f,\pi^M)
	\right| \le \frac{M_T}{M}c_2 
\right\} \ge \mathbb{P}\left\{	
	\frac{
		2 \|f\|_\infty \|g\|_\infty M_T
	}{
		M (g,q^M)
	}  \le \frac{M_T}{M}c_2 
\right\}
\label{eqCharranga-L2}
\end{equation}
and if we choose 
\begin{equation*}
c_2 = \frac{
	2\|f\|_\infty\|g\|_\infty 
}{
	\left(
		1 - \frac{1}{\sqrt{2}}
	\right) (g,q)
},
\end{equation*}
then
\begin{eqnarray}
\mathbb{P}\left\{	
	\frac{
		2 \|f\|_\infty \|g\|_\infty M_T
	}{
		M (g,q^M)
	}  \le \frac{M_T}{M}c_2 
\right\} &=& \mathbb{P}\left\{
	(g,q^M) \ge \left(
		1 - \frac{1}{\sqrt{2}}
	\right) (g,q) 
\right\} \nonumber\\
&=&  \mathbb{P}\left\{
	M(g,q^M) - M(g,q) \ge - \frac{M}{\sqrt{2}} (g,q) 
\right\}.
\label{eqCharranga-L2-2}
\end{eqnarray}

Since $(g,q)=E_q[(g,q^M)]$ and $(g,q^M)$ is the sum of $M$ independent, and bounded, random variables taking values within the interval $(0,\|g\|_\infty]$ (recall that $g>0$), we can readily apply Hoeffding's tail inequality \cite{Hoeffding63} on Eq. (\ref{eqCharranga-L2-2}) to obtain
\begin{equation}
\mathbb{P}\left\{
	M(g,q^M) - M(g,q) \ge - \frac{M}{\sqrt{2}} (g,q) 
\right\} \ge 1 - \exp\left\{
	-\frac{(g,q)^2}{\|g\|_\infty^2}M
\right\}.
\label{eqAhuevo-L2}
\end{equation}
Substituting (\ref{eqAhuevo-L2}) back into (\ref{eqCharranga-L2-2}) and (\ref{eqCharranga-L2}) yields the desired result,
\begin{equation*}
\mathbb{P}\left\{
	\left|
		(f,\check \pi^M) - (f,\pi^M)
	\right| \le \frac{M_T}{M}c_2 
\right\} \ge 1 - \exp\left\{
	-c_2' M
\right\},
\end{equation*}
where $c_2' = \frac{(g,q)^2}{\|g\|_\infty^2} > 0$.

\section{Proof of Lemma \ref{L3}}
\label{App_3}

The first part of Lemma \ref{L3} follows from the combination of Lemmas \ref{L1} and \ref{L2}. We first note that, from Lemma \ref{L1},
\begin{equation}
\mathbb{P}\left\{
	\left| (f,\bar \pi^M) - (f,\check \pi^M) \right| > c_1\frac{M_T}{M}
\right\} < \exp\{ -c_1'M\}
\label{eqKK1}
\end{equation}
for sufficiently large $M$, while Lemma \ref{L2} implies 
\begin{equation}
\mathbb{P}\left\{
	\left| (f,\check \pi^M) - (f, \pi^M) \right| > c_2\frac{M_T}{M}
\right\} < \exp\{ -c_2'M\},
\label{eqKK2}
\end{equation}
where $c_2=c_2'=(g,q)^2 / \| g\|_\infty^2$. Let $c=c_1+c_2$. Then, since 
\begin{equation*}
| (f,\bar \pi^M) - (f,\pi^M) | \le | (f,\bar \pi^M) - (f,\check \pi^M) | + | (f,\check \pi^M) - (f,\pi^M) |,
\end{equation*}
we trivially obtain that
\begin{equation}
\mathbb{P}\left\{
	\left| (f,\bar \pi^M) - (f,\pi^M) \right| > c\frac{M_T}{M}
\right\} \le \mathbb{P}\left\{
	\left| (f,\bar \pi^M) - (f,\check \pi^M) \right| +
	\left| (f,\check \pi^M) - (f,\pi^M) \right| > c\frac{M_T}{M}
\right\}.\\
\label{eqLink1}
\end{equation}
However, if 
$$
\left| (f,\bar \pi^M) - (f,\check \pi^M) \right| +
	\left| (f,\check \pi^M) - (f,\pi^M) \right| > c\frac{M_T}{M}
$$
is true, then
$$ 
\left| (f,\bar \pi^M) - (f,\check \pi^M) \right| > c_1\frac{M_T}{M} 
\quad \mbox{or} \quad 
\left| (f,\check \pi^M) - (f,\pi^M) \right| > c_2\frac{M_T}{M},
$$
or both jointly, are true. Therefore, 
\begin{eqnarray}
\mathbb{P}\left\{
	\left| (f,\bar \pi^M) - (f,\check \pi^M) \right| +
	\left| (f,\check \pi^M) - (f,\pi^M) \right| > c\frac{M_T}{M}
\right\} && \nonumber \\
\le \mathbb{P}\left\{
	\left| (f,\bar \pi^M) - (f,\check \pi^M) \right| > c_1\frac{M_T}{M} 
\right\} && \nonumber \\
+ \mathbb{P}\left\{
	\left| (f,\check \pi^M) - (f,\pi^M) \right| > c_2\frac{M_T}{M} 
\right\} &&\nonumber \\
\le \exp\left\{-c_1'M \right\} + \exp\left\{-c_2'M\right\} &&\nonumber \\
= 2\exp\left\{-\frac{(g,q)^2}{\| g \|_\infty^2} M \right\},&&
\label{eqLink2}
\end{eqnarray}
for sufficiently large $M$, where the second inequality follows from (\ref{eqKK1}) and (\ref{eqKK2}), and the equality is due to the fact that $c_1'=c_2'$.

Combining (\ref{eqLink1}) and (\ref{eqLink2}) yields the first part of Lemma \ref{L3}, with $c' = (g,q)^2 / \| g \|_\infty^2$.

The second part of Lemma \ref{L3} follows from a standard Borel-Cantelli argument. Indeed, let ${\mathcal E}_M$ be the event in which $\left| 
		(f,\bar \pi^M) - (f, \pi^M)
	\right| > c\frac{
		M_T 
	}{
		M
	}
$ holds true. From the first part of the Lemma,
\begin{equation*}
\mathbb{P}\left\{
	{\mathcal E}_M
\right\} < 2\exp\left\{
	-c' M
\right\},
\end{equation*}
with $c'>0$, for sufficiently large $M$ (specifically, for all $M > M_a$, with $M_a$ as in the proof of Lemma \ref{L1}). Therefore,
\begin{equation*}
\sum_{M=1}^\infty \mathbb{P}\left\{
	{\mathcal E}_M
\right\} \le M_a + \sum_{M=M_a+1}^\infty \exp\left\{
	-c' M
\right\} < \infty,
\end{equation*}
because $M_a<\infty$ and $\sum_{M=M_a+1}^\infty \exp\left\{
	-c' M
\right\} < \infty$. As a consequence (see, e.g., \cite[Theorem 2.7]{Williams91}), 
$
\mathbb{P}\left\{ 
	\lim\sup {\mathcal E}_M
\right\} = 0,
$
which implies that 
\begin{equation*}
\lim_{M\rightarrow\infty} | (f,\bar \pi^M) - (f, \pi^M) | = 0 \quad \mbox{a.s.} 
\end{equation*}
\qed


\bibliographystyle{plain}         
\bibliography{biblio}


\end{document}